\documentclass[aps,prc,twocolumn,showpacs,superscriptaddress]{revtex4-1}

\usepackage{amsfonts,amsthm,latexsym}
\usepackage{graphicx}
\usepackage{bm}
\usepackage{xcolor}
\usepackage{soul} 
\usepackage{slashed}
\usepackage{mathrsfs}
\usepackage{latexsym}
\usepackage{longtable}
\usepackage{graphicx}
\usepackage{epsfig}
\usepackage{natbib}
\usepackage{amsmath}
\usepackage{bbm}
\usepackage{amssymb}
\usepackage{color}
\usepackage{hyperref}
\usepackage{url}
\usepackage{multirow}
\usepackage[utf8]{inputenc}
\usepackage{soul}
\usepackage{subfigure}
\usepackage{scalerel}

\begin{document}

\title{Delta baryons and diquark formation in the cores of neutron stars}

\author{Germ\'an Malfatti} \email{gmalfatti@fcaglp.unlp.edu.ar}
\affiliation{Grupo de Gravitaci\'on, Astrof\'isica y
  Cosmolog\'ia,Facultad de Ciencias Astron\'omicas y Geof\'isicas,
  Universidad Nacional de La Plata, Paseo del Bosque S/N, La Plata
  (1900), Argentina.}  \affiliation{CONICET, Godoy Cruz 2290, Buenos
  Aires (1425), Argentina.}
 \author{Milva G.  Orsaria}%
\email{morsaria@fcaglp.unlp.edu.ar} \affiliation{Grupo de
  Gravitaci\'on, Astrof\'isica y Cosmolog\'ia,Facultad de Ciencias
  Astron\'omicas y Geof\'isicas, Universidad Nacional de La Plata,
  Paseo del Bosque S/N, La Plata (1900), Argentina.}
\affiliation{CONICET, Godoy Cruz 2290, Buenos Aires (1425),
  Argentina.}  \author{Ignacio F. Ranea-Sandoval}%
\email{iranea@fcaglp.unlp.edu.ar} \affiliation{Grupo de Gravitaci\'on,
  Astrof\'isica y Cosmolog\'ia,Facultad de Ciencias Astron\'omicas y
  Geof\'isicas, Universidad Nacional de La Plata, Paseo del Bosque
  S/N, La Plata (1900), Argentina.}  \affiliation{CONICET, Godoy Cruz
  2290, Buenos Aires (1425), Argentina.}  \affiliation{Department of
  Physics, San Diego State University, 5500 Campanile Drive, San
  Diego, CA 92182, USA.}
 
 \author{Gustavo A. Contrera}
 \affiliation{CONICET, Godoy Cruz 2290, Buenos Aires (1425),
   Argentina.}  \affiliation{IFLP, UNLP, CONICET, Facultad de Ciencias
   Exactas, Diagonal 113 entre 63 y 64, La Plata (1900), Argentina.}
 
 \author{Fridolin Weber}
 \affiliation{Department of Physics, San Diego State University, 5500
   Campanile Drive, San Diego, CA 92182, USA.}  \affiliation{Center
   for Astrophysics and Space Sciences, University of California, San
   Diego, La Jolla, CA 92093, USA}

\date{\today}

\begin{abstract}
We investigate the hadron-quark phase transition in cold neutron stars
in light of (i) the observed limits on the maximum-mass of heavy
pulsars, (ii) constraints on the tidal properties inferred from the
gravitational waves emitted in binary neutron-star mergers, and (iii)
mass and radius constraints derived from the observation of hot spots
on neutron star observed with NICER. Special attention is directed to the possible
presence of $\Delta(1232)$ baryons in neutron star matter.  Our
results indicate that this particle could make up a large fraction of
the baryons in neutron stars and thus have a significant effect on the
properties of such objects, particularly on their radii. This is
partially caused by the low density appearance of $\Delta$s for a wide
range of theoretically defensible sets of meson--hyperon, SU(3) ESC08
model, and meson--$\Delta$ coupling constants.  The transition of
hadronic matter to quark matter, treated in the 2SC+s condensation
phase, is found to occur only in neutron stars very close to the mass
peak.  Nevertheless, quark matter may still constitute an appreciable
fraction of the stars' total matter if the phase transition is treated
as  Maxwell-like (sharp), in which case the neutron stars
located beyond the gravitational mass peak would remain stable against
gravitational collapse. In this case, the instability against
gravitational collapse is shifted to a new (terminal) mass different
from the maximum-mass of the stellar sequence, giving rise to stable
compact objects with the same gravitational masses as those of the
neutron stars on the traditional branch, but whose radii are smaller
by up to 1 km. All models for the equation of state of our study fall
comfortably within the bound established very recently by Annala {\it
  et al.} (Nature Physics, 2020)
\end{abstract}

\maketitle


\section{\label{sec:level1}Introduction}

The observations of $2 \, M_\odot$ binary pulsars PSR J1614-2230
\cite{Demorest:2010bx}, PSR J0348+0432 \cite{Antoniadis:2013pzd}, PSR
J2215+5135 \cite{Linares2018}, and PSR J0740+6620
\cite{2020NatAs...4...72C} strongly constrains theoretical models of
the equation of state (EoS) of ultra-dense nuclear matter (see, for
example,
Refs.\ \cite{annurev-nucl-102711-095018,NSMmax,annurev-astro-081915-023322},
and references therein). Moreover, the analysis of data from the
binary neutron star (BNS) merger events GW170817
\cite{GW170817-detection} and GW190425 \cite{gw190425-detection} and
from the Neutron Star Interior Composition Explorer (NICER) instrument
\cite{Riley2019,Raaijmakers2019,Bilous2019,Miller2019,Bogdanov2019,%
  Bogdanov2019a,Guillot2019} made it possible to put additional, tight
constraints on the behavior of matter at densities higher than nuclear
saturation density, $n _0$.

One of the most important conclusions obtained from the data of
  GW170817 is that the radius of a $1.4 \, M_\odot$ neutron star (NS)
  is constrained to $R_{1.4} < 13.6$ km (see, for example,
  Ref.\ \cite{Raithel2018}). Moreover, based on the data of GW170817,
  it has been argued that a NS could not support a mass larger than
  $M_{\rm NS}^{\rm max} \sim 2.3 \, M_\odot$
  \cite{NSMmax}. Considering this additional constraint it follows
  that $R_{1.4} = 11.0^{+0.9}_{-0.6} \, M_\odot$
  \cite{Capano2020sco}. An improved analysis of the GW170817 data
has restricted the originally determined tidal deformability $\Lambda
_{1.4} < 800$ of this NS to $\Lambda _{1.4}= 190^{+390}_{-160}$
\cite{GW170817-new}.

The second BNS merger, GW190425, was detected on April 25th, 2019 with
the LIGO Livingston interferometer.  To date, an electromagnetic
counterpart associated with this event has not been detected. The
inferred mass of the primary object is, under the low-spin (high-spin)
assumption, $M_1 = 1.60 - 1.87 \, M_\odot$ $(M_1 = 1.61 - 2.52
M_\odot)$, and for the secondary object $M_2 = 1.46 - 1.69 \, M_\odot$
$(M_2 = 1.12 - 1.68 M_\odot)$. With a total gravitational mass of
$M_{\rm tot} = 3.4^{+0.4}_{-0.1}\, M_\odot$, this is the most massive
BNS system ever detected, differing by five standard deviations from
the Galactic BNS mean value of $\sim 2.69 \, M_\odot$ (see, for
example, Ref.\ \cite{BNS-mass}). The fact that the signal of GW190425
was only detected by one interferometer and that no electromagnetic
counterpart has been observed renders the constraints on the mass and
radius of this NS not as tight as those obtained with
GW170817. Nevertheless, there are indications that a massive ($M > 1.7
\, M_\odot$) NS would have a radius larger that $R\sim 11$ km
\cite{gw190425-detection}.

Observations of the isolated pulsar PSR J0030+0451 made with the NICER
instrument produced two independent measurements of the pulsar's mass
and radius, $M=1.34^{+0.15}_{-0.14} \, M_\odot$ and an equatorial
radius of $R_{\rm eq} = 12.71^{+1.14}_{-1.19}$ km \cite{Riley2019},
and $M=1.44^{+0.15}_{-0.14} \, M_\odot$ and $R_{\rm eq} =
13.02^{+1.24}_{-1.06}$ km \cite{Miller2019}.

Last but not least we mention the very recent work \cite{Landry2020}
where the limits on the maximum NS mass, gravitational-wave data, and
information about neutron star masses and radii from X-ray emissions
have been used to arrive at $R_{1.4} = 12.32^{+1.09}_{-1.47}$ km for
the radius of a $1.4 \, M_\odot$ NS.

Both the existence of $\sim 2 M_{\odot}$ pulsars as well as the data
from gravitational-wave events of BNS mergers suggest that the NS EoS
needs to be relatively soft at low and intermediate nuclear densities
in order to achieve relatively small radii for $\sim 1.4 \, M_\odot$
NSs, such as those quoted above, but much stiffer at high densities to
accommodate heavy $2 \, M_\odot$ NSs too.  One possible theoretical
scenario leading to such a behavior of the EoS is obtained if NS
matter undergoes a phase transition from hadronic matter to deconfined
quark matter \cite{Baym:2017whm,Orsaria:2019ftf, Alford_2019JPG,
    PhysRevLett.122.061101,Annala2020}. Neutron star models
containing such matter are referred to as hybrid stars (HSs). At low
and intermediate nuclear densities, the matter in the cores of such
stars is assumed to be composed of neutrons, protons, and hyperons,
while at higher densities these particles give way to the formation of
quark matter. made of deconfined up ($u$), down ($d$), and strange
($s$) quarks. The transition of one phase of matter to the other is
generally modeled as a Maxwell transition or a Gibbs transition \cite{
  Bhattacharyya_2010,PhysRevC.96.025802,PhysRevD.99.083014}.
Depending on the hadron-quark surface tension \cite{MARUYAMA2008192,
  PhysRevC.88.045803}, the transition region is characterized
either by a jump from one phase to the other (Maxwell case), or the
existence of a mixed phase where pressure varies smoothly with density
(Gibbs case).

If quark matter exists in the interiors of NSs, it ought be in a color
superconducting state
\cite{alford-annurev.nucl.51.101701.132449,shovkovy-2005,alford-RevModPhys.80.1455}. Such
a state would be energetically favored, since a system of weakly
interacting fermions at low temperatures is unstable with respect to
the formation of diquarks, similarly to the formation of Cooper pairs
in ordinary superconductors.  (For recent studies of quark matter in
NSs, see \cite{Malfatti2019hot, PhysRevC.101.055204,
    PhysRevLett.124.171103,Annala2020}, and references therein.)  One
possible condensation pattern of color superconducting quark matter,
which is studied in this paper, is the so-called 2SC+s phase
\cite{alford-annurev.nucl.51.101701.132449,PhysRevC.96.065807}, which
is expected to occur when the strange quark is too massive to
participate in the formation of pairs with $u$ and $d$ quarks. In this
case, only green and red $u$ and $d$ quarks can form diquark
condensates due the symmetry breaking of the SU(3)$_{\rm color}$
group.

The possible existence of hyperons in the cores of NSs has been
investigated by numerous authors using either phenomenological or
microscopic approaches for the neutron star matter EoS with hyperons
(see Ref.\ \cite{glendenning2012compact,Vidana2016} for comprehensive
lists of references). Depending on the microscopic many-body theory,
it has been found that such particles may appear rather abundantly in
NS matter at densities just a few times higher than the nuclear
    saturation density $n_0$ \cite{KATAYAMA201543,
  PhysRevC.95.065803}. The situation is different for the charged
states of the $\Delta$ baryons. In fact, the possible presence of this
particle in NSs has long been ignored because early studies carried
out with the relativistic mean-field theory suggested that $\Delta$s
would only appear at densities greater than $\sim 10 \, n_0$, too high to
be reached in the cores of NSs \cite{1985ApJGlen}. Updated microscopic
models and tighter constraints on the model parameters, however, paint
a different picture
\cite{Weber1989,PhysRevC.94.045803,Spinella2020:WSBook,LI2018234,
  PhysRevC.90.065809, KOLOMEITSEV2017106, Ribes_2019}.  These studies
show that $\Delta$s could in fact make up a large fraction of the
baryons in neutron star matter and thus have a significant effect on
the properties of NSs. In particular, the radii of NSs are sensitive
to the $\Delta$ population \cite{schurhoff2010neutron,
  PhysRevC.92.015802, PhysRevC.94.045803}. The relevance of $\Delta$s
for heavy ion collisions and different nuclear physics processes has
been emphasized in
\cite{Waldhauser1988.PRC,Waldhauser1987,FERREIRA2002303}.

In this work, we investigate the hadron-quark phase transition in cold neutron stars
in light of the observed limits on the maximum-mass of heavy pulsars,
constraints on the tidal properties inferred from the gravitational
waves emitted in binary neutron-star mergers, and mass and radius
constraints derived from the observation of hot spots on neutron
star observed with NICER. 
The details
of the construction of the hybrid EoS as well as the equilibrium and
charge neutrality conditions are given in Sect.\ \ref{heos}. For the
description of the hadronic matter, presented in
Sect.\ \ref{sec:level2}, we use a density dependent relativistic
mean-field model which includes the strange mesons $\sigma^*$ and
$\phi$. All members of the baryon octet as well as the $\Delta$ baryons
are included in our model. In Sect.\ \ref{sec:level3}, we provide the
details of the non-local quark model used to describe the quark phase
inside of HSs, including the possibility of 2SC+s color
superconductivity. Section \ref{sec:level4} is devoted to the
presentation and discussion of the results. The conclusions are given
in Sect.\ \ref{sec:level5}. Finally, details of the 2SC+s phase
calculations are provided in Appendix \ref{append}.

\section{The hybrid EoS} \label{heos}

We model the matter in the inner cores of NSs under the hypothesis of
a hadron-quark phase transition. We use the SW4L parametrization to
model the matter at low nuclear densities and use a non-local chiral
quark model to describe the matter at high nuclear densities. For the
construction of the corresponding hybrid EoS, there are some general
characteristics and considerations to be taken into account, as
discussed below.

The phase transition of hadronic to quark matter is modeled by using
both the Maxwell and the Gibbs formalism.  The systematics of the
phase transition is intimately related to the unknown value of the
hadron-quark surface tension, $\sigma _{\rm HQ}$. If this value is
greater than a critical value, estimated to be around 70 MeV/fm${}^2$,
a sharp phase transition will be favored, where matter changes from
hadronic matter to pure quark matter at a certain radial location
inside a HS \cite{criticalSurface:2003,criticalSurface:2014}. This
situation is described by the Maxwell formalism. For this type of
phase transition, the pressure is isobaric in the transition region
and the EoS is characterized by an energy gap at the interface between
hadronic and quark matter.  In this scenario, the electric chemical
potential might not always be continuous along the interface (for a
more detailed discussion, see Ref. \cite{glendenning2012compact}).

On the other hand, if $\sigma _{\rm HQ}$ is lower that the critical
value, the favored scenario is the one in which a mixed phase is
formed where hadrons and quarks coexists. This type of phase
transition is described by the bulk Gibbs formalism, where the
electric charge is conserved globally. For intermediate cases of
$\sigma _{\rm HQ}$, where one has to take into account both Coulomb
and surface energy contributions, a series of geometrical structures
(blobs, rods, and slabs), also called the pasta phase, might appear
(see \cite{Orsaria:2019ftf,pasta:2019}, and references therein). The
nature and characteristics of this phase are strongly dependent on the
value of $\sigma _{\rm HQ}$.

The hope is that NS data will help to shed light on the
possible hadron-quark phase transition in the inner cores of
NSs. Neutron star masses and radii are generally considered to be
possible the primary indicators, but clues may be provide by other
pointers as well.  One such pointer could be the {\it speed} at which
the conversion of hadronic matter to quark matter proceeds. As it has
been shown recently \cite{lugones-extended}, if the phase transition
is sharp and the conversion rate {\it slow} (with respect to the
characteristic oscillation frequency time-scale), then compact stars
located beyond the gravitational mass peak will remain stable. In this
case, the instability against gravitational collapse is then shifted
to a new {\it terminal} mass different from the maximum-mass of a
compact-star sequence. On the contrary, if the conversion is {\it
  fast}, the traditional stability criteria for stellar configurations
against radial oscillations is recovered. This phenomena could give
rise to a new family of twin-like stars, stars with the same
gravitational masses as ordinary compact stars but different
radii. The standard twin-like stars scenario has been studied for several
different hybrid EoSs
\cite{benic-twins,Ranea:2016,PhysRevLett.119.161104}.

\subsection{Equilibrium conditions}\label{ssec:equilcond}

Equilibrium conditions for the hybrid EoS implies thermal, chemical
and mechanical equilibrium. Since we are considering cold hybrid
matter, thermal equilibrium between the hadronic and quark phase is
automatically satisfied.

Chemical equilibrium of nucleons, hyperons and quarks in the cores of
hybrid stars depends not only on the chemical reactions occurring
between them, but also on the local density. For the low nuclear
density phase, we consider the chemical equilibrium given by
\begin{equation}
\mu_B = \mu_n + q_B\,\mu_e \, ,
\label{chempot}
\end{equation}
where $q_B$ is the baryon electric charge and $\mu_n$ and $\mu_e$ are
the neutron and electron chemical potentials, respectively.

In the case of quark matter, we need to deal with quark flavors and
quark colors, which, in principle, should lead to six different
chemical potentials. In particular, the presence of color
superconductivity breaks down the color gauge symmetry SU(3)$_{\rm
  color}$ into the subgroups U(1)$_3$ and U(1)$_8$ leading to two
independent chemical potentials, $\mu_3$ and $\mu_8$ respectively,
associated with the color charges. In the 2SC+s phase, strange quark
decouples from the superconducting system of up and down quarks (see
Appendix \ref{append} for details). Red and green quarks are
degenerate, and diquarks condense in the blue direction, as it happens
for two-flavor color superconductors (2SC) \cite{huang2003charge,
  PhysRevD.73.114019}. Thus, we can take $\mu_3=0$ so that $\mu_8$
remains as the only chemical potential related to the color
charges. Therefore, chemical equilibrium of the quark phase is given
by
\begin{eqnarray}
 \mu_{ur} &=& \mu_{ug} = \mu - \frac{2}{3}\mu_e + \frac{1}{2\sqrt{3}} \mu_8 \, , \nonumber \\
 \mu_{ub} &=&  \mu - \frac{2}{3}\mu_e - \frac{1}{\sqrt{3}} \mu_8 \, , \nonumber \\
 \mu_{dr} &=& \mu_{dg} = \mu + \frac{1}{3}\mu_e + \frac{1}{2\sqrt{3}} \mu_8 \, , \nonumber \\
 \mu_{db} &=&  \mu + \frac{1}{3}\mu_e - \frac{1}{\sqrt{3}} \mu_8 \, , \nonumber \\
 \mu_{sr} &=& \mu_{sg} = \mu_{dr} \, , \nonumber \\
 \mu_{sb} &=& \mu_{db} \, ,
\end{eqnarray}
where $\mu \equiv \mu_n/3$.

{{Electrons and muons satisfy the condition
\begin{eqnarray}
 \nu_{\mu} + \bar\nu_e  + e^- \leftrightarrow \mu^- \, ,
\end{eqnarray}
which implies for the chemical potentials of these particles
\begin{eqnarray}
\mu_{\mu}=\mu_e + \mu_{\bar\nu_e}+ \mu_{{\nu}_{\mu}} \, .
\end{eqnarray}
For cold NSs, as considered in this work, the neutrino chemical
potentials are zero and $\mu_{\mu} = \mu_e$.  }}

Mechanical equilibrium
of hybrid matter is guaranteed through the condition
\begin{eqnarray}
 P^H(\mu_B^H, \mu_e^H, \{\alpha _{j}\} ) = P^q(\mu^q, \mu_e^q,
 \{\kappa _{k}\} ) \, ,
\label{eq:GibbsP}
\end{eqnarray}
where the quantities $\{ \alpha _{j}\}$ and $\{ \kappa _{k}\}$ in
Eq.\ (\ref{eq:GibbsP}) represents the field variables characterizing
the solutions to the field equations of the hadronic and quark phases,
respectively.  As it was mentioned before, due the uncertainty of the
surface tension $\sigma _{\rm HQ}$, one has to assume a priori the
nature of the first-order phase transition, to be either sharp
(Maxwell-like) or smooth (Gibbs-like). For both cases, the transition
from the low density (hadronic phase) to the high density (quark
phase) is possible as long as the Gibbs free energy of the quark phase
is lower than the Gibbs free energy of the hadronic phase. The Gibbs
free energy, at zero temperature, is given by
\begin{equation}
 G_{E} = \sum_i \frac{\mu_i n_i}{n_B} \, ,
 \label{gibssenergy}
\end{equation}
where $n_B$ is the baryon number density and $\mu_i$ denotes the
chemical potential of each particle species $i$ present in the
system. The quantity
\begin{equation}
n_i=-\frac{\partial\Omega}{\partial\mu_i} \, ,
\label{densit_part}
\end{equation}
represents the number density of a particle of type $i$, which is
obtained from the corresponding thermodynamic potential (see
Sect.\ \ref{ssec:chargeneut} for the leptonic contributions and
Sects.\ \ref{sec:level2} and \ref{sec:level3} for details related to
the hadronic and quark phases, respectively).  Once the grand
canonical potential of the system is obtained, the pressure is
obtained from $P = -\Omega$ and the energy density of the system
follows from
\begin{equation}
 \epsilon = - P + \sum_i \mu_i \, n_i \, .
 \label{eq:EoS}
\end{equation}

Assuming a sharp  Maxwell phase transition, the condition of
chemical equilibrium given by $ G_{E}^H = G_{E}^q$ must be satisfied
together with Eq.\ (\ref{eq:GibbsP}). In this case, there is a jump in
the energy density between the hadronic and quark phases and the
pressure is constant during the transition.

In the case of a smooth Gibbs phase transition, a mixed phase of
hadrons and quarks is formed and the pressure grows monotonically
in the transition region. Therefore, not only Eq.\ (\ref{eq:GibbsP})
must be taken into account, but the following equations
\begin{eqnarray}
n_{B}^{\mathrm{mix}} &=& (1-\chi) n_B^H + \chi n_B^q \, ,\nonumber\\
\epsilon^{\mathrm{mix}} &=& (1-\chi) \epsilon^H + \chi\epsilon^q\, ,
\end{eqnarray}
are to be taken into account as well. Here $n_B^H$ ($\epsilon^H$) and
$n_B^q$ ($\epsilon^q$) are the baryon number (energy) densities of
each phase. The quantity $\chi \equiv V_q/V$ denotes the volume
proportion of quark matter, $V_q$, in the unknown volume
$V$. Therefore, $0 \le \chi \le 1$ by definition, depending on how
much hadronic matter has been converted into quark matter
\cite{glendenning2012compact}.

\subsection{Charge neutrality condition}\label{ssec:chargeneut}

In addition to the pressure condition given by Eq.\ (\ref{eq:GibbsP}),
one needs to impose on the field equations either local or global electric
and color charge neutrality, depending on the nature of the phase
transition. For a Maxwell transition, the local electric charge
conservation reads
\begin{equation}
\sum_{i,l} q_{i,l}^{H(q)}\,n_{i,l}^{H(q)} = 0 \, ,
\end{equation}
where $q_i$ is the electric charge of all particles in the hadronic
($H$) or quark ($q$) phases. The quantity $q_l$ is the corresponding
expression for the electric charges of leptons. The particle number
densities $n_{i,l}$ are obtained by making use of
Eq.\ (\ref{densit_part}) for each type of particle.

Regarding the color charge neutrality condition, it is known that
strange quark matter is color neutral. However, for the 2SC+s phase,
due the SU(3)$_{\rm color}$ symmetry breaking, diquarks are not color
neutral. Thus, we require
\begin{equation}
\frac{\partial\Omega}{\partial\mu_8} = \frac{1}{\sqrt{3}}\left(n_r +
n_g -2 n_b \right)= 0 \, ,
\end{equation}
where $r$, $g$, $b$ stand for red, green, and blue colors,
respectively. Note that the condition $\mu_3=0$, mentioned in
Sect.\ \ref{ssec:equilcond}, implies that $n_r=n_g$.

In the case of a Gibbs phase transition, the condition of global
electric charge neutrality is given by
\begin{equation}
\sum_{i,l}\left[(1 - \chi) q_{i,l}^H \, n_{i,l}^H + \chi q_{i,l}^q \,
  n_{i,l}^q \right] = 0 \, .
\end{equation}
In contrast to local electric charge neutrality, the global charge
neutrality condition allows for a positive net electric charge in the
hadronic phase, which makes matter more isospin symmetric, and a net
negative electric charge in the quark matter phase. In  other words, the
concept of global charge conservation involves only the mixed phase but not the
pure hadronic matter phase  or pure quark matter phase.

In this work we consider, for both the hadronic and quark phases, that
the leptonic contribution comes from electrons and muons treated as 
free Dirac particles. The thermodynamic potential is thus given by
\begin{eqnarray}
\Omega_l &=& -\frac{1}{\pi^2}\sum_B \int^{p_{F_l}}_0 \! dp \,
\frac{p^4}{\sqrt{p^2+m_l^{2}}},
\label{omegalep}
\end{eqnarray}
where $p_{F_l}$ are the Fermi momenta of leptons of mass $m_l$.  We
use $m_e= 0.5$ MeV and $m_\mu =105.66$ MeV.

\section{The Hadronic Phase}
\label{sec:level2}

For the description of hadronic matter, we use the density dependent
nonlinear relativistic mean-field model with the SW4L parametrization
\cite{Spinella:2018dab,Spinella2020:WSBook}.
{\bf{This model accounts for medium effects by making the
    meson-baryon coupling of the $\rho$-meson dependent on the local
    baryon number density.  Models that consider a density dependence
    for all mesons have first been introduced in
    \cite{Typel:1999yq}. The $\rho$-meson coupling used in our paper
    has the same density dependence as the one in
    \cite{Typel:1999yq}.}}
    One of the
    advantages of this model is that by choosing proper Gaussian or
    Lorentzian functional forms for the density dependence, the
    slope of the symmetry energy  can be fixed without affecting other nuclear
    properties or the stiffness of the nuclear EoS. The
    slope of the symmetry energy has become very important for NS matter
    calculations due to
    its effect on the composition and properties of neutron stars
    \cite{Spinella2017:thesis}.

The interactions among the baryons are
described by the exchange of $\sigma$, $\omega$, $\rho$, $\sigma^*$,
and $\phi$ mesons.  The lagrangian of this model is given by
\begin{eqnarray}
  \mathcal{L} &=& \sum\limits_B \overline\psi_B\bigl[\gamma_{\mu}
    (i\partial^{\mu}-g_{\omega B}\omega^{\mu}-g_{\phi
      B}\phi^{\mu}-\tfrac{1}{2}g_{\rho B}(n)\boldsymbol{\tau}\cdot
    \boldsymbol{\rho}^{\mu})\nonumber\\ &&-(m_B-g_{\sigma
      B}\sigma-g_{\sigma^*
      B}\sigma^*)\bigr]\psi_B\nonumber\\
  &&+\tfrac{1}{2}\left(\partial_{\mu}\sigma\partial^{\mu}\sigma
  -m^2_{\sigma}\sigma^2\right)\nonumber\\ &&-\tfrac{1}{3}b_{\sigma}m_n\left(g_{\sigma
    N}\sigma\right)^3 -\tfrac{1}{4}c_{\sigma}\left(g_{\sigma
    N}\sigma\right)^4\nonumber\\ &&-\tfrac{1}{4}\omega_{\mu\nu}\omega^{\mu\nu}
  +\tfrac{1}{2}m^2_{\omega}\omega_{\mu}\omega^{\mu}\nonumber\\ &&
  -\tfrac{1}{4}\boldsymbol{\rho}_{\mu\nu}\cdot\boldsymbol{\rho}^{\mu\nu}\nonumber
  +\tfrac{1}{2}m^2_{\rho}\boldsymbol{\rho}_{\mu}\cdot\boldsymbol{\rho}^{\mu}\\
  &&-\tfrac{1}{4}\phi^{\mu\nu}\phi_{\mu\nu}+\tfrac{1}{2}m^2_{\phi}\phi_{\mu}
  \phi^{\mu}\nonumber\\ &&+\tfrac{1}{2}\left(\partial_{\mu}
  \sigma^*\partial^{\mu}\sigma^* -m^2_{\sigma^*}\sigma^{*2}\right)\, ,
 \label{laghad}
\end{eqnarray}
where the sum over $B$ is over all the baryons in the baryon octet as
well as the four electrically charged states of the $\Delta$
resonance. Baryon-baryon interactions are modeled in terms of scalar
($\sigma, ~\sigma^*$), vector ($\omega,~ \phi$), and isovector
($\rho$) meson fields. The quantities $g_{\rho B}(n)$ denote density
dependent {{isovector}} meson--baryon coupling constants given by
\begin{equation}
g_{\rho B}(n) = g_{\rho B}(n_0)\,\mathrm{exp}\left[\,-a_{\rho}
  \left(\frac{n}{n_0} - 1\right)\,\right] \, ,
\end{equation}
where $n = \sum_B n_B$ is the total baryon number density. Once the
field equations for the baryons and mesons are obtained by solving the
equations of motion that follow from Eq.\ (\ref{laghad}), we use the
relativistic mean-field approximation in which the meson field
operators are replaced by their mean-field values. By virtue of this
procedure, we obtain a coupled, nonlinear algebraic system of meson
mean-field equations,
\begin{eqnarray}
m_{\sigma}^2 \bar{\sigma} &=& \sum_{B} g_{\sigma B} n_B^s -
\tilde{b}_{\sigma} \, m_N\,g_{\sigma N} (g_{\sigma N} \bar{\sigma})^2
\nonumber\\ & & - \tilde{c}_{\sigma} \, g_{\sigma N} \, (g_{\sigma N}
\bar{\sigma})^3 \, \nonumber\\ m_{\sigma^*}^2 \bar{\sigma^*} &=&
\sum_{B} g_{\sigma^* B} n_B^s\, , \nonumber\\ m_{\omega}^2 \bar{\omega}
&=& \sum_{B} g_{\omega B} n_{B}\, , \\ m_{\rho}^2\bar{\rho} &=&
\sum_{B}g_{\rho B}(n)I_{3B} n_{B} \, , \nonumber\\ m_{\phi}^2
\bar{\phi} &=& \sum_{B} g_{\phi B} n_{B}\, , \nonumber
\end{eqnarray}
where $I_{3B}$ is the 3-component of isospin, and $n_{B}^s$ and
$n_{B}$ are the scalar and particle number densities for each baryon
$B$, which are given by
\begin{eqnarray}
n_{B}^s&=& \frac{1}{4\pi^2} \int^{p_{F_B}}_0 \frac{d^3p}{(2 \pi)^3}
\frac{m_B^*}{\sqrt{p^2+m_B^{*2}}}, \\ n_{B}&=& \frac{p_{F_B}^3}{ 3 \pi^2 }
\, .
\end{eqnarray}
Here $p_{F_B}$ is the Fermi momentum and $m_B^*= m_B - g_{\sigma
  B}\bar{\sigma}-g_{\sigma^* B}\bar{\sigma^*}$ is the effective baryon
mass.

\begin{figure}[htb] 
\begin{center}
\includegraphics[width=0.48\textwidth]{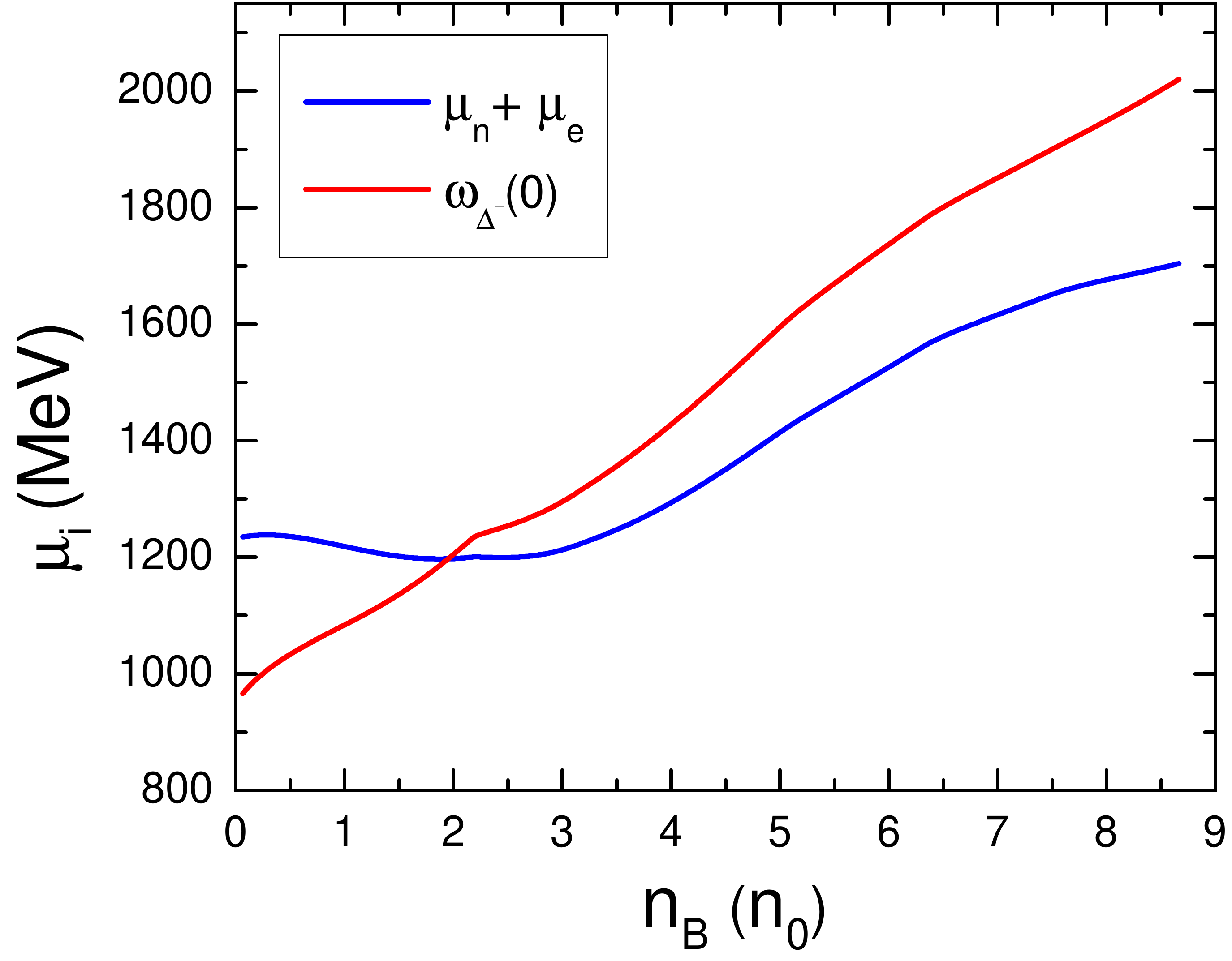}
\caption{(Color online) {\bf{Comparison of the
        neutron-plus-electron effective chemical potential,
        $\mu_n+\mu_e$, with the lowest single-particle energy state of
        the $\Delta^-$, $\omega_{\Delta^-}(0)$, in NS matter. The
        presence of $\Delta^-$ particles is triggered at the density 
        where the two curves cross, at around $2 n_0$.}}}
 \label{chem_pot_delta}
\end{center}
\end{figure}

{\bf{ The chemical equilibrium condition $\mu_n +\mu_e = \mu_B$
      of NS matter was already defined in Eq.\ (\ref{chempot}). Since
      $\mu_B = \omega_B(p_{F_B})$, where $\omega_B(p_{F_B})$ is the
      single-baryon energy,
\begin{eqnarray}
\omega_B(p) &=& g_{\omega B} \bar{\omega} + g_{\rho B}(n) \bar{\rho}
I_{3B}\nonumber \\ &+&g_{\phi B} \bar{\phi} +\sqrt{p^2_{F_B}+ 
    m_B^{*2} } + \widetilde{R} \, , 
\label{eq:omegaB}
\end{eqnarray}
at the Fermi surface, the $\Delta^-$ state becomes populated in NS
matter once the density is high enough so that $\mu_n +\mu_e =
\omega_{\Delta^-}(0)$ is fulfilled. The situation is graphically
illustrated in Fig.\ \ref{chem_pot_delta} where we compare the density
dependences of $\mu_n+\mu_e$ and $\omega_{\Delta^-}(0)$ with each
other, computed for the hadronic model (SWL4) of this work. As can
been seen from this figure, equality between $\mu_n+\mu_e$ and
$\omega_{\Delta^-}(0)$ is already reached at densities of just around
twice nuclear saturation density (we will come back to this issue in
Sect.\ \ref{sec:level4} (see Fig.\ \ref{Pops_Hyb}) there), which are
easily reached in the cores of neutron stars.}}

{\bf{The term $\widetilde{R} = [\partial g_{\rho B}(n)/\partial n]
    I_{3B} n_B \bar{\rho}$ in Eq.\ (\ref{eq:omegaB}) is the
    rearrangement term necessary to guarantee thermodynamic
    consistency \cite{Hofmann:2001}.}} This term also affects the
    pressure of hadronic matter, which is given by
\begin{eqnarray}
P_h &=& \frac{1}{\pi^2}\sum_B \int^{p_{F_B}}_0 \! dp \, 
\frac{p^4}{\sqrt{p^2+m_B^{*2}}}-\frac{1}{2} m_{\sigma}^2
\bar{\sigma}^2\nonumber\\ &-& \frac{1}{2} m_{\sigma^*}^2
\bar{\sigma^*}^2 + \frac{1}{2} m_{\omega}^2 \bar{\omega}^2 +
\frac{1}{2} m_{\rho}^2 \bar{\rho}^2+ \frac{1}{2} m_{\phi}^2
\bar{\phi}^2\\ &-& \frac{1}{3} \tilde{b}_{\sigma} m_N (g_{\sigma N}
\bar{\sigma})^3 - \frac{1}{4} \tilde{c}_{\sigma} (g_{\sigma N}
\bar{\sigma})^4 + n \widetilde{R}.  \nonumber
\label{HM:pressure}
\end{eqnarray}

\begin{table}[htb]
\begin{center}
\begin{tabular}{|c|c|}
\hline 
$~~$Quantity$~~$ & $~~$SW4L Parameters$~~$\\ \hline
$m_{\sigma}$  (GeV)    & 0.5500          \\
$m_{\omega}$  (GeV)          &0.7826    \\
$m_{\rho}$  (GeV)          & 0.7753         \\
$m_{\sigma^*}$  (GeV)          & 0.9900         \\
$m_{\phi}$  (GeV)          & 1.0195         \\
$g_{\sigma N}$             & 9.8100         \\
$g_{\omega N}$            & 10.3906           \\
$g_{\rho N}$            & 7.8184           \\
$g_{\sigma^* N}$            & 1.0000           \\
$g_{\phi N}$            & 1.0000           \\
$\tilde{b}_{\sigma}$         &0.0041                     \\
$\tilde{c}_{\sigma}$         &$- 0.0038$                \\
$a_{\rho}$         &0.4703            \\ \hline
\end{tabular}
  \caption{Parameters of the SW4L parametrization that lead to the
    properties of symmetric nuclear matter at saturation density given
    in Table \ref{table:properties}.}
\label{table:parametrizations}
\end{center}
\end{table}

\begin{table}[htb]
\begin{center}
\begin{tabular}{|c|c|}
\hline 
$~~$Saturation Properties$~~$ & $~~$SW4L$~~$\\ \hline
$n_0$  (fm$^{-3}$)     & 0.150          \\
$E_0$  (MeV)          & -16.0     \\
$K_0$  (MeV)          & 250.0          \\
$ m^*_N/m_N$             & 0.7        \\
$J_0$  (MeV)          & 30.3           \\
$L_0$  (MeV)          & 46.5         \\ \hline
\end{tabular}
  \caption{Properties of nuclear matter at saturation density, $n_0$,
    obtained for the SW4L parametrization. The entries are: energy per
    nucleon $E_0$, nuclear incompressibility $K_0$, effective nucleon
    mass $m^*$, symmetry energy $J_0$, and slope of the symmetry energy
    $L_0$.}
\label{table:properties}
\end{center}
\end{table}
In this work we use the parameter set SW4L shown in Table
\ref{table:parametrizations}. The coupling constants as well as the
parameters $\tilde{b}_{\sigma}$, $\tilde{c}_{\sigma}$, and $a_{\rho}$
were adjusted according to the properties of nuclear matter at
saturation density listed in Table \ref{table:properties}.

{
The scalar meson-hyperon coupling constants $g_{\sigma Y}$ and
$g_{\sigma^* Y}$ are fit to hyperon ($Y$) single-particle potentials
and self-potentials derived from the available empirical data on
hypernuclei, once the vector meson-hyperon couplings $g_{\omega Y}$ and
$g_{\phi Y}$ have been specified.  In SU(3) symmetry the vector
couplings are given in terms of the vector mixing angle $\theta_V$,
the vector coupling ratio $\alpha_V$, and the meson singlet-to-octet
coupling ratio $z$ \cite{Dover1984} (see also
\cite{Schaffner1996,Spinella2017:thesis,Spinella2020:WSBook,Spinella:2018dab}).
The values of these parameters are $\theta_V=37.50^{\circ}$,
$\alpha_V=1$, and $z=0.1949$, corresponding to the SU(3) ESC08 model
\cite{ECSO_SU3}.

Once the vector meson-hyperon coupling constants are specified, the
scalar meson-hyperon couplings are set to reproduce empirical hyperon
single-particle potentials in symmetric nuclear matter at nuclear
saturation, $U_{Y}^{~\!(N)}(n_0)$,
using the relation
\cite{Spinella:2018dab}
\begin{equation} \label{eq:hyperon-nucleon-potential}
  U_{Y}^{~\!(N)}(n_0) = g_{\omega Y} \bar\omega + g_{\phi Y} \bar\phi
  - g_{\sigma Y} \bar\sigma \, .
\end{equation}
The following hyperon potentials have been employed:
$U_{\Lambda}^{(N)}(n_0) = -28$ MeV, $U_{\Sigma}^{(N)}(n_0) = +30$ MeV,
and $U_{\Xi}^{(N)}(n_0) = -14$ MeV.
The strange-scalar meson-$\Lambda$ coupling constant
$g_{\sigma^*\Lambda}$ has been set to reproduce a saturation
self-potential of $U_{\Lambda}^{(\Lambda)}(n_0)=-1$ MeV in
isospin-symmetric $\Lambda$-matter, {\bf a value close to that}}
suggested by the Nagara event \cite{Ahn2013}, using the following,
    \begin{equation} \label{eq:lambda-lambda-potential}
  U_{\Lambda}^{~\!(\Lambda)}(n_0) = 
	  g_{\omega \Lambda} \bar\omega_0
  + g_{\phi \Lambda} \bar\phi_0
  - g_{\sigma \Lambda} \bar\sigma_0
  - g_{\sigma^* \Lambda} \bar\sigma^*_0\,. \nonumber
\end{equation}
{\bf{From this event, the $\Lambda\Lambda$ binding energy was
    originally determined to be $1.01\pm 0.20$ MeV. This value has
    subsequently been revised to $0.67\pm 0.17$ MeV \cite{Ahn2013} due
    to the change of the $\Xi^-$ mass by the particle data group.
    Both values consistently suggest a weak attractive $\Lambda\Lambda$
    interaction. We note that values of $U_{\Lambda}^{(\Lambda)}(n_0)=
    -1$ or $-5$ MeV have been employed in the literature in the past,
    while phenomenological relativistic mean-field approaches suggest
    values between approximately $-14$ and $+9$ MeV, depending on how
    tight SU(6) constraints are imposed on the approaches
    \cite{PhysRevC.95.065803}.}}

The other strange-scalar meson-hyperon couplings are determined
relative to that of the $\Lambda$ using $U_{\Xi}^{(\Xi)}(n_0) = 2
U_{\Lambda}^{(\Lambda)}(n_0/2)$, so that $g_{\sigma^*\Sigma} =
g_{\sigma^*\Lambda} = 1.9242$ \cite{Oertel2015}.  The isovector-vector
meson-hyperon coupling constants $g_{\rho Y}$ are given by
$g_{\rho\Lambda} = 0$ and $g_{\rho\Sigma} = g_{\rho\Xi} = g_{\rho
  N}$.
  
To adjust the SW4L parametrization to the nuclear properties of Table
\ref{table:properties} we define $x_{\phi B} = g_{\phi B} / g_{\omega
  N}$. With this definition, the $\phi$--$Y$ coupling ratios are
given by $x_{\phi\Lambda} = x_{\phi\Sigma} = 1.7855$ and $x_{\phi\Xi}
= 7.7247$.

Most studies of $\Delta$s in dense matter have been conducted in the
standard relativistic mean-field (RMF) approach
\cite{Waldhauser1987,Waldhauser1988.PRC,Weber1989.JPG,Choudhury1993.PRC,%
  PhysRevC.92.015802,Lavagno2010}, the density-dependent RMF approach
\cite{PhysRevC.90.065809,KOLOMEITSEV2017106,Spinella2020:WSBook}, or
the (density-dependent) relativistic Hartree-Fock approach
\cite{Weber1989,PhysRevC.94.045803}, all indicating at the abundant
existence of $\Delta$s in NS matter. We note, however, that a recent
study performed for the quark-meson coupling model has suggested that
$\Delta$ isobars are absent in NSs \cite{motta2020delta}. The reason
for that are the many-body forces generated by the change in the
internal quark structure of the baryons in the scalar mean fields
generated in dense nuclear matter.
  
All of these studies suffer from the problem that the meson--$\Delta$
couplings are only poorly constrained so that particular coupling sets
must be chosen with which to conduct the analysis. The meson--$\Delta$
coupling space has been systematically investigated in
Ref.\ \cite{Spinella2020:WSBook,LI2018234} and will be further explored and
constrained in this work.

  To
include $\Delta$s in the study of dense NSs matter, we follow a
two-pronged approach. First we shall consider a quasi-universal
meson--$\Delta$ coupling scheme
\begin{equation}
x_{{\sigma} {\Delta}} = x_{{\omega} {\Delta}} = 1.1\, ~~  x_{{\rho}
  {\Delta}} =x_{{\phi} {\Delta}} =1.0\, ~~ x_{{\sigma^*} {\Delta}} =0.0\, ,
\end{equation}
{{where $x_{\sigma^* B} \equiv g_{\sigma^* B}/g_{\sigma^* \Lambda}$
  and $g_{\sigma^* \Lambda} = 1.9242$}}. Next, we explore the parameter
space of the $\sigma$--$\Delta$ coupling constant, related with the
effective $\Delta$ mass, $m_{\Delta}^*$, considering the constraint
imposed by the event GW170817 on NSs radii. We study coupling ratios
in the interval $1.1 \leq x_{\sigma \Delta} \leq 1.258$.  The lower
bound $x_{\sigma\Delta} = 1.1$ (together with $x_{\omega\Delta}=1.1$,
$x_{\rho\Delta}=1.0$) satisfied the constraints on the potential of
$\Delta$s in symmetric nuclear matter at saturation density
\cite{Spinella2020:WSBook,PhysRevC.90.065809,KOLOMEITSEV2017106,Riek2009}.
The upper bound $x_{\sigma\Delta} =1.258$ is determined by the
microscopic stability of matter, that is, for values $x_{\sigma\Delta}
> 1.258$ pressure is no longer monotonously increasing with density so
that the matter becomes microscopically unstable.

\section{The Quark Phase}
\label{sec:level3}

For the description of quark matter, including diquarks in the SU(3)
non-local model, we use the Lagrangian given by
\begin{eqnarray}\label{lagrangiano}
\mathcal{L} &=& \bar \psi(x)(-i \slashed{\partial} + \hat{m} )\psi(x)
- \frac{G_s}{2}\big[j_{a}^s(x)j_{a}^s(x) + \nonumber \\ &+&
  j_{a}^p(x)j_{a}^p(x)\big] +
\frac{G_v}{2}j_{a}^{\mu}(x)j_{a}^{\mu}(x) \nonumber \\ &-&
\frac{H}{4}A_{abc}\big[j_{a}^s(x)j_{b}^s(x)j_{c}^s(x) -
  3j_{a}^s(x)j_{b}^p(x)j_{c}^p(x) \big] \,  ,
  \label{lagnonl}
\end{eqnarray}
where $A_{abc}$ are constants given in terms of the Gell-Mann matrices
and $j_{a}^s(x)$, $j_{a}^p(x)$ and $ j_{a}^{\mu}(x)$ are interaction
currents. The current quarks masses and the coupling constants $G_s$,
$H$, and $\Lambda$ are taken from Ref.\ \cite{Malfatti2019hot}. The
vector interaction coupling constant, $G_v$, is expressed in terms of
the scalar coupling constant, $G_s$, and is treated as a free
parameter.

To include diquark channels in the model, an additional term of the
form
\begin{equation}
\mathcal{L}_D= -\frac{G_D}{2} \left[j_D \left(x\right) \right]^\dagger
j_D \left(x\right) \, ,
\end{equation}
needs to be added to Eq.\ (\ref{lagnonl}). Here $G_D$ is the diquark
coupling constant expressed in multiples of $G_s$. The diquarks currents
are given by
\begin{equation}
j_D\left(x\right) = \int d^4\,  z g(z) \overline{\psi}_C \left(x +
\frac{z}{2}\right)i\gamma^5 \lambda_A \lambda_{A'}\psi\left( x -
\frac{z}{2}\right) \, ,
\end{equation}
where $\psi_C = \gamma_2 \gamma_4 \overline{\psi}^T(x)$.  The matrices
$\lambda_A$ and $\lambda_{A'}$ operate in the color and flavor spaces,
respectively, and take on values ${2,5,7}$ (see Appendix \ref{append}
for details). The non-local regulator $g(x-y)$ is related to its
momentum space representation, $g(p)$, via
\begin{equation}
g(x - y) = \int \frac{d^4p}{(2\,\pi)^4} \, e^{i (x-y) p} \, g(p) \, .
\end{equation}

The inclusion of color superconductivity leads to a matrix for the
diquark condensates that can be written as
\begin{equation}
 s_{AA'}= \langle \overline{\psi}_C \gamma_5\lambda_A \lambda_{A'}
 \psi\rangle \, ,
\end{equation}
where $C = \gamma_2 \gamma_4$ is the operator of charge
conjugation. This matrix can be simplified by a color rotation,
\begin{equation}
s = 
 \begin{pmatrix}
s_{22} & 0 & 0 \\
s_{52} & s_{55} & 0 \\
s_{72} & s_{75} & s_{77} \\
\end{pmatrix}  \,  .
\end{equation}
The non-diagonal matrix components are negligible in the one-gluon
exchange regime \cite{Alford:1998mk,Fritzsch:1973pi}, so that one only
needs to keep the elements $s_{22}$ , $s_{55}$, and $s_{77}$.  In this
work, we consider the formation of $(u_r,d_g)$ and $(u_g,d_r)$
diquark pairs.  Therefore
\begin{eqnarray}
s_{22}= \langle \overline{\psi} \gamma_5\lambda_2 \lambda_2'
\psi\rangle\, , \quad s_{55} = s_{77} = 0 \, .
\end{eqnarray}

Including the new diquark bosonic field $\bar\Delta$ and its
associated auxiliary field $\bar D$, we bosonize the euclidean
effective action, $S_E$, which follows from
Eq.\ (\ref{lagnonl}). Then, in the mean-field approximation,
\begin{eqnarray}
\frac{S_E^{MFA}}{V^{(4)}}&=&-2\,
\mathrm{Tr}\,\int\,\frac{d^4p}{(2\,\pi)^4}\,\mathrm{ln}\,A(p)\,-
\nonumber \\ &-&\frac{1}{2}\left[ \left(\bar\sigma_a \bar S_a + \frac{G_s}{2}\bar
  S_a\,\bar S_a + \bar\theta_a \bar V_a - \frac{G_v}{2}\bar V_a\,\bar
  V_a\right) \right. \nonumber \\ &+& \left. \frac{H}{2}A_{abc} \bar
  S_a \bar S_b\bar S_c + 2 \bar \Delta \bar D +G_D \bar D\,\bar D
  \right] \, , 
\label{action}
\end{eqnarray}
where $A(p)$ is the inverse of the quark propagator with interactions
and $\bar S_a$ and $\bar V_a$ are the mean-field values of the
auxiliary fields corresponding to $\bar \sigma_a$ and $\bar \theta_a$,
respectively.  After some algebra in the first term of
Eq.\ (\ref{action}) (see Appendix \ref{append} for details) the
regularized thermodynamic potential for the 2SC+s phase reads
\begin{eqnarray} 
 \Omega = &-& 2\sum_c \int\frac{d^4p}{(2\pi)^4} \left\{
 \mathrm{ln}\left[\frac{{q_{sc}^+}^2 + M_{sc}^2}{p_{sc}^2 +
     m_s^2}\right]- \frac{1}{2} \mathrm{ln} |A_c|^2 \right.\nonumber
 \\ &-& \left.  \mathrm{ln}[({p_{uc}^+}^2 + m_u^2 ) ({p_{dc}^+}^2 +
   m_d^2)] \right\} \nonumber \\ &-& \frac{1}{2} \sum_f \left[
   \left(\bar \sigma_f \bar S_f + \frac{G_S}{2} \bar S_f^2 + \bar
   \theta_f \bar V_f - \frac{G_V}{2} \bar V_f^2\right)
   \right. \nonumber \\ &+& \left. \frac{H}{2} \bar S_u \bar S_d \bar
   S_s + 2\bar \Delta \bar D +G_D \bar D^2\right]
 -\Omega_{free}^{Reg}\, .
 \label{Omegon}
 \end{eqnarray}
All the quantities in Eq.\ (\ref{Omegon}) including the expression for
$\Omega_{free}^{Reg}$ are given in Sect.\ \ref{append}. From the following set
of (seven) coupled equations,
\begin{equation}
\frac{\partial \Omega}{\partial \bar\sigma_f} =
0\, , \quad \frac{\partial \Omega}{\partial \bar\theta_f} =
0\, , \quad \frac{\partial \Omega}{\partial \bar\Delta} = 0 \, ,
\end{equation}
it is possible to determine the mean-field values of $\bar\sigma_f$,
$\bar\theta_f$, and $\bar\Delta$.

\section{\label{sec:level4} Results and Discussion}

\begin{figure}[htb]
\begin{center}
\includegraphics[width=0.49\textwidth]{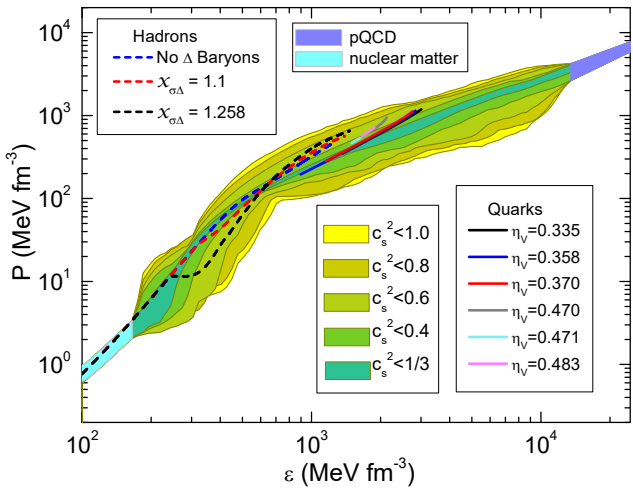}
\caption{(Color online) Comparison of the EoSs of this work with
  bounds on the EoS recently established in Ref.\ \cite{Annala2020}. }
 \label{fig:eos}
\end{center}
\end{figure}

In Fig.\ \ref{fig:eos} we compare the EoSs of this work with bounds on
the EoS recently established in Ref.\ \cite{Annala2020}.  The bands
cover a large number of EoSs generated with the speed-of-sound
interpolation method.
As can be seen, all our models lie comfortably within the bounds on
the neutron star matter EoS shown in Fig.\ \ref{fig:eos}.

\begin{figure}[b]
\begin{center}
\includegraphics[width=0.49\textwidth]{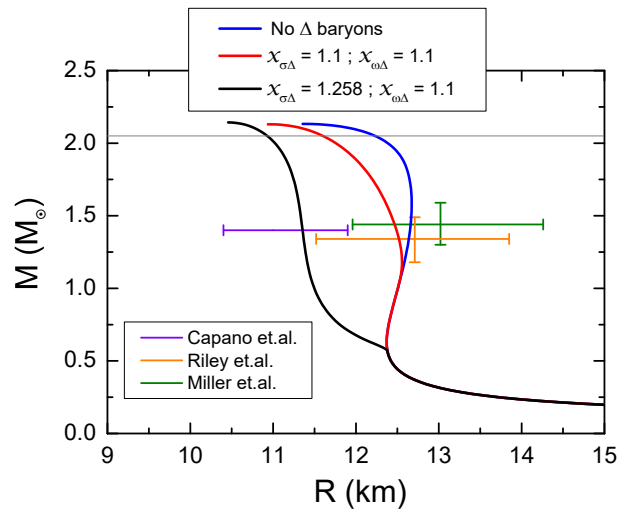}
\caption{(Color online) Mass-radius relationships of compact stars
  computed for purely hadronic matter, based on the SW4L
  parametrization introduced in Sect.\ \ref{sec:level2}. The observed
  radius constraints are taken from Refs.\ \cite{Riley2019,
    Miller2019} (orange and green horizontal lines, respectively) and
  Ref.\ \cite{Capano2020sco} (purple horizontal line). The gray
  horizontal line shows the minimum gravitational mass established
  for PSR J0740+6620 \cite{2020NatAs...4...72C}.}
 \label{MR_hads}
\end{center}
\end{figure}

In Fig.\ \ref{MR_hads}, we show the mass-radius relationships of
stellar configurations, computed for the hadronic SW4L EoS of this
work, for different values of the $\sigma$--$\Delta$ coupling ratio
$x_{\sigma \Delta}$. All the three mass-radius relationships obey the
gravitational-mass constraint set by $2 \, M_\odot$ pulsars as well as
the radius constraints extracted from NICER observations
    \cite{Riley2019, Miller2019} and the gravitational-wave event
    GW170817 \cite{Capano2020sco}.
As can be seen in Fig.\ \ref{MR_hads}, the impact of $\Delta$ baryons
on the mass-radius relationship is strong for $\sigma$--$\Delta$
coupling ratios in the range of $1.1 \leq x_{\sigma\Delta} \leq
1.258$. {{The mass-radius relationships of this figure are computed
    for a $\omega$--$\Delta$ coupling constant ratio of
    $x_{\omega\Delta} = 1.1$. We found that if we set
    $x_{\sigma\Delta}$ = $x_{\omega\Delta}$, the minimum coupling
    value for the EoS to remain microscopically stable ($c_s >0$)
    is $x_{\sigma\Delta}$ = 1.1. Varying $x_{\sigma\Delta}$ while
    keeping $x_{\omega\Delta}$ at 1.1 changes the maximum NS mass
    insignificantly, but the radii of all stars decrease if
    $x_{\sigma\Delta} > x_{\omega\Delta}$ and increase
    significantly if $x_{\sigma\Delta} < x_{\omega\Delta}$,
      where $x_{\sigma\Delta} = 1.258$ sets the
      upper limit for which $c_s>0$ holds for
    $x_{\omega\Delta} = 1.1$. Finally, the
    mass-radius relationships are virtually the same if
    $x_{\sigma\Delta} = x_{\omega\Delta}$ regardless whether
    $x_{\sigma\Delta} = 1.1$ or 1.258.  As can
      be seen in Fig.\ \ref{MR_hads}, current observational constrains
      on the radius of a $\sim 1.5 \, M_\odot$ NS are reproduced by
      our models for $1.1 \leq x_{\sigma\Delta} \leq 1.258$. The hope
      is that future observational constraints will allow one to
      narrow down this range and to draw firm conclusions on the
      possible existence of $\Delta$s in NSs.}  } 

In summary, we note that the presence of $\Delta$s in NS matter
strongly modifies the radii of NSs. The $2\, M_\odot$ mass constraint
can nevertheless be fulfilled comfortably.  This is in agreement with
what has been found in other studies \cite{schurhoff2010neutron,
  LI2018234, KOLOMEITSEV2017106, Ribes_2019}.  
\begin{figure}[tb]
\begin{center}
\includegraphics[width=0.49\textwidth]{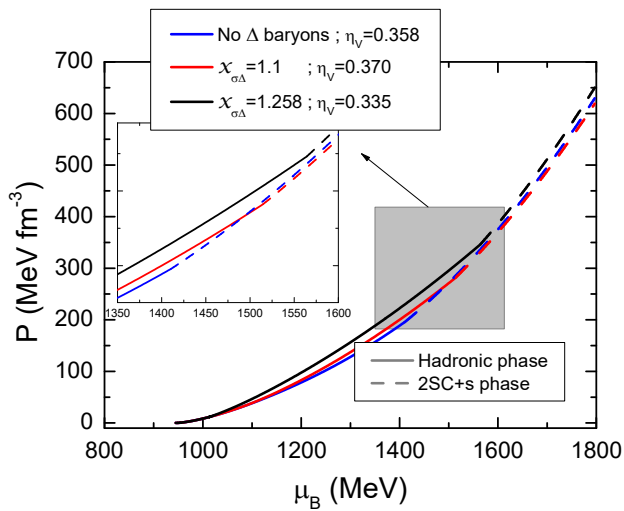}
\caption{(Color online) Illustration of the hybrid EoSs computed in
  this work. Quark matter is treated as a 2SC+s color superconductor.
  The inset figure shows the pressures of the different phases of
  matter in the phase transition region.}
 \label{EoS_hyb}
\end{center}
\end{figure}
The radii of NSs with canonical masses between 1.4 to $1.5\, M_\odot$
turn out to be particularly sensitive to the presence of
$\Delta$s. They may change by up to $\sim 1.5$ km for the
theoretically defensible sets of meson--hyperon (SU(3) ESC08 model) and
meson--$\Delta$ coupling constants of this work.

\begin{figure}[htb]
\begin{center}
\includegraphics[width=0.49\textwidth]{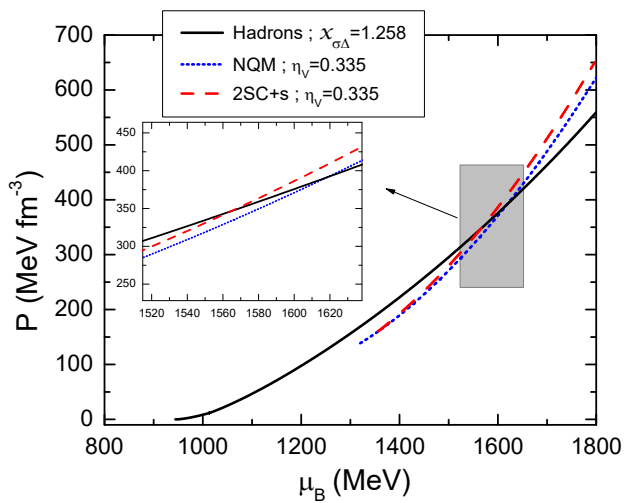}
\caption{(Color online) Comparison of hybrid EoSs. NQM refers to
  ordinary non-superconducting quark matter, 2SC refers to quark
  matter in the 2SC+s superconducting phase. The latter turns out to
  be energetically favored at high chemical potentials. The inset
  figure shows the pressures in the phase transition region.}
 \label{NQM_2SC}
\end{center}
\end{figure}

In Fig.\ \ref{EoS_hyb}, we show the results for the hybrid EoS
computed for the models introduced in Sects.\ \ref{sec:level2} and
\ref{sec:level3}.  The solid lines mark the region where matter exists
solely in the hadronic matter phase and the dashed lines mark the
region where the matter exists in the form of quark matter in the
2SC+s color superconducting phase.  Also shown in this figure is the
impact of $\Delta$ baryons on the EoS, which depends on the $x_{\sigma
  \Delta}$ coupling ratio as discussed just above, and the role of the
quark vector interaction value, $\eta_v$ ($\equiv G_v/G_s$), whose
value determines the pressure at which the hadron-quark phase
transition takes place. {{The 2SC+s phase is always energetically
    favored relative to normal ({\it{i.e}}., non-superconducting)
    quark matter, as shown in Fig.\ \ref{NQM_2SC}. Moreover, this
    result is independent of the vector interaction value
    considered.}}  We therefore
find that a direct transition from hadronic matter to 2SC+s color
superconducting quark matter for our model, bypassing ordinary quark
matter.  The inset figure shows the pressures of the different phase
of matter in the phase transition zone. We can see that when the
$x_{\sigma \Delta}$ coupling constant ratio is increased, the transition
pressure increases as well. Moreover, larger values of the vector
interaction lead to a stiffer EoS.

\begin{figure}[htb]
\begin{center}
\includegraphics[width=0.49\textwidth]{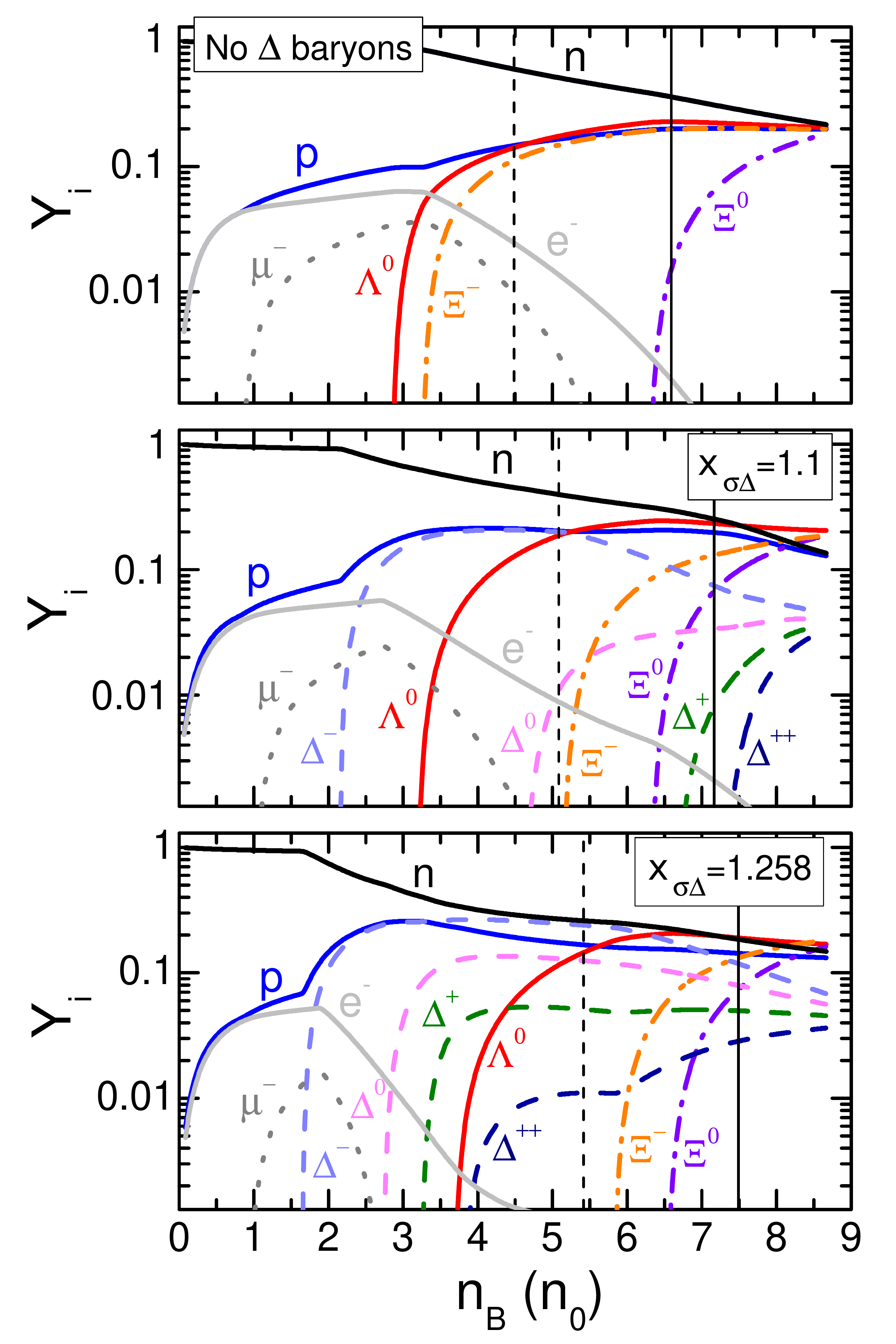}
\caption{(Color online) Baryon-lepton populations of the neutron stars
  shown in Fig.\ \ref{MR_hads}. The solid vertical lines mark the
  central densities of the maximum-mass stars associated with these
  compositions. The dashed vertical lines mark the densities at which
  phase equilibrium with 2SC+s quark matter would set in.}
 \label{Pops_Hyb}
\end{center}
\end{figure}

In Fig.\ \ref{Pops_Hyb}, we show the particle populations of neutron
star matter computed for the hadronic EoSs of this work. In the top
panel we show how the composition looks like if the $\Delta$ baryon
is not taken into account in the calculation.  The other two panels
show the hadronic populations if all states of the $\Delta$ baryon
($\Delta^{++}$, $\Delta^{+}$, $\Delta^{0}$, $\Delta^{-}$) are taken
into account in the calculation.  Naively, one would assume $\Delta$s
would not be favored in NS matter for several reasons
\cite{Spinella2020:WSBook}. First, their rest mass is greater than the
rest masses of both the $\Lambda$ and $\Sigma$ hyperons. Second,
negatively charged baryons are generally favored as their presence
reduces the high Fermi momenta of the leptons, but the $\Delta^-$ has
triple the negative isospin of the neutron ($I_{3\Delta^-} = -3/2$),
and thus its presence should be accompanied by a substantial increase
in the isospin asymmetry of the system. These arguments, however
appear to be largely invalid for the following reasons.  Incorporating
the repulsive saturation potential of the $\Sigma$ hyperon into the
determination of the meson--$\Sigma$ coupling constants greatly
reduces the $\Sigma$'s favorability (it is totally absent in the
compositions shown in Fig.\ \ref{Pops_Hyb}), and thus it is not likely
to compete with the $\Delta^-$ state. More importantly the overall
effect of the asymmetry energy on the system is significantly reduced
when one employs a parametrization with a density-dependent isovector
meson--baryon coupling constant as done in this work (SW4L), which is
necessary to satisfy the constraints on the slope of the asymmetry
energy at saturation density \cite{Spinella2020:WSBook}.

The values of the $\sigma$--$\Delta$ coupling ratio are
$x_{\sigma \Delta}=1.1$ and $x_{\sigma \Delta}=1.258$. As a reminder,
the latter value constitutes the maximum possible value allowed by
microscopic stability of the matter.  We see that the appearance of
the charged states of the $\Delta$ baryon is sequential, beginning
with the $\Delta^-$ at less than twice nuclear saturation density and
ending with the $\Delta^{++}$ at densities as low as around 4 times
nuclear saturation density, depending on the value of $x_{\sigma
  \Delta}$.
Based on these populations, $\Delta$ baryons are abundantly present
in NS matter already at densities that are markedly smaller
than the densities of the maximum-mass neutron stars (solid vertical
lines) associated with these compositions. Even NSs with a
masses in the range between 1.4 to $1.5 \, M_\odot$ would possess
significant populations of $\Delta$s, which, as was shown in
Fig.\ \ref{MR_hads}, significantly modifies the radii of these
objects.  We also note that the $\Delta$ population sets in at
densities that are less than the density at which the hadron-quark
phase transition would set in (vertical dashed lines in
Fig.\ \ref{Pops_Hyb}) for this parametrization.  This is most evident
for the maximum possible value of the $\sigma$--$\Delta$ coupling
ratio, $x_{\sigma \Delta}=1.258$, in which case all charged $\Delta$
states are present well before the threshold density
at which quark deconfinement sets in.  It is interesting to note that,
at over certain density ranges, the $\Delta^-$ abundances are
comparable to those of protons and $\Lambda$s.  Given the impact
$\Delta$s may have on the masses and radii of NSs, one
might hope that future astrophysical observations of these and other
NS quantities (e.g., moment of inertia) will help to
elucidate the relevance of $\Delta$s for dense nuclear matter
studies.

\begin{figure}[htb]
\begin{center}
\includegraphics[width=0.49\textwidth]{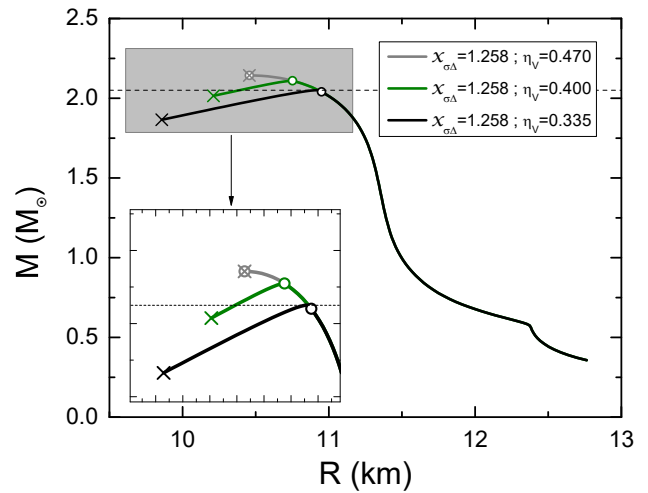}
\caption{Mass-radius relationships obtained with the hybrid EoS for
  different vector repulsion parametrizations. The hollow circles
  indicate the onset of quark matter and the crosses mark the
  terminal mass model of each  stellar sequence.}
 \label{Fig_new}
\end{center}
\end{figure}

\subsection{Extended branch of stable hybrid stars}\label{ssec:extended_branch}

{The confinement/deconfinement process is not solely ruled by
  the strong interaction, whose timescale is $\sim 10^{-23}$
  s. Other physical phenomena like Coulomb screening and surface and
  curvature effects play important roles (see
  Ref.\ \cite{Lugones-Grunfeld2017}, and references
  therein). Moreover, it is important to stress that the strong
  interaction operates on time-scales that are shorter (by several
  orders of magnitude) than those related to the weak
  interactions. For this reason, the weak
    interaction cannot operate during the
  deconfinement process. In view of that, newly deconfined
  quark matter is transitorily out of chemical equilibrium and the abundances per baryon of each particle
  need to be the same in both phases. Several model-dependent
  calculations show that if quark matter is to be produced preserving
  flavor, its final equilibrium state is not accessible directly and a
  two-step transition between hadronic and quark matter must take
  place, firstly to a flavor preserving out of $\beta$-equilibrium
    quark state, followed by a second weak decay to the final
  equilibrium quark state in $\sim 10^{-8}$ s (see, for example,
  Ref.\ \cite{Lugones2016}, and references therein).}

{ Since there is no high-density EoS constructed from
  first-principles, it is not clear whether a fluid element that
  oscillates around the transition pressure will suffer a slow or
  rapid direct conversion. Several works have shown that the
  probability of a hadron-quark phase transition is related to a
  model-dependent timescale (see, for example,
  Refs.\ \cite{Lugones-Grunfeld2011,Bombacietal2016,Lugones2016}, and
  references therein). In addition, there are some results that
  indicate this timescale is around $\sim 10^{-3}$ s
  \cite{Haenseletal1989} or even larger (see, for example,
  Refs.\ \cite{Iida-Sato1998,
    Bombaci-Parenti-Vidana2004,Bombacietal2009}). These theoretical
  studies indicate that the hadron-quark phase transition should be
  considered to be slow.}
{{Because of  these theoretical uncertainties, we consider both the
  slow and rapid conversion scenarios between hadronic and quark
  matter and analyze their astrophysical implications.}}

In Fig.\ \ref{Fig_new}, we show the mass-radius relationship of HSs
for a fixed value of $x_{\sigma\Delta}$ but different values of the
vector repulsion parameter $\eta_v$. The onset of quark matter in the
cores of these stars is marked with hollow circles. It can be seen
that quark deconfinement occurs only for stars very close to the
maximum-mass peak of each stellar sequence. This is in agreement with
results reported in the literature previously (see, for example,
\begin{figure}[htb]
\begin{center}
\includegraphics[width=0.48\textwidth]{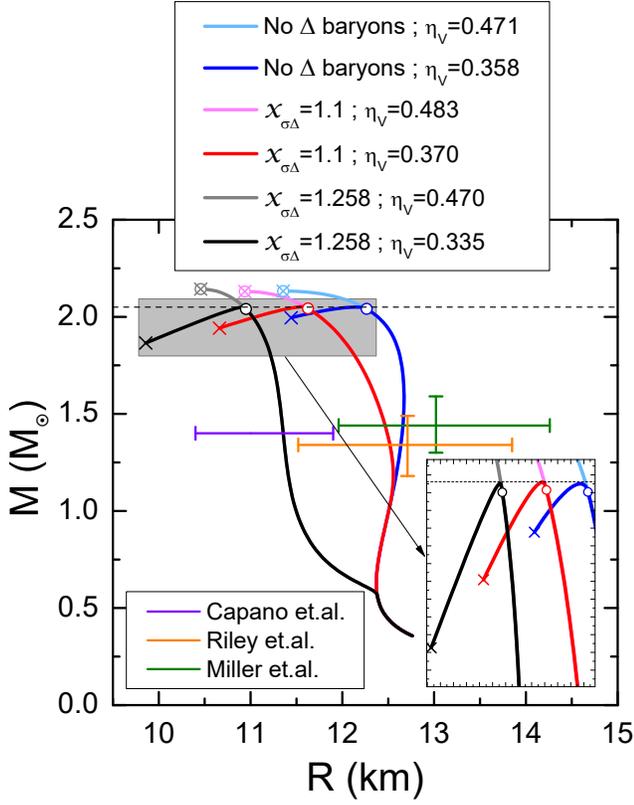}
\caption{(Color online) Mass-radius relationships obtained with the
  hybrid EoS of this work ($x_{\omega\Delta}=1.1$). The constraints on
  $M$ and $R$ are the same as in Fig.\ \ref{MR_hads}. The hollow
  circles mark the onset of quark deconfinement. The extended branches
  of stable HSs terminate at the crossed locations.}
 \label{MR_hyb}
\end{center}
\end{figure}

Refs.\ \cite{Ranea:2016,Malfatti2019hot, mariani2019}, and references
therein), where it was shown that the {\it rapid} conversion of
hadronic matter to quark matter in HSs tends to destabilize such
  objects.  For a rapid conversion, the timescale associated with
  transforming hadronic matter to quark matter in a star is much shorter than the timescale set by the stellar perturbations
  oscillations \cite{lugones-extended, mariani2019}.  The situation is
  dramatically different if the conversion proceeds {\it slowly}, that
  is, if the timescale associated with transforming hadronic matter to
  quark matter is much larger than the timescale set by the stellar
  perturbations. In the latter case, a new (extended) branch of
  stable HSs can exist, ranging from the maximum-mass star of a
  sequence to a new terminal-mass configuration
  \cite{lugones-extended, mariani2019}.  They are marked with crosses
  in Fig.\ \ref{Fig_new}. As shown in Refs.\ \cite{lugones-extended,
    mariani2019} the usual static stability condition against
  gravitational collapse, $\partial M/\partial\epsilon_c \geq 0$
  (where $\epsilon_c$ is the central energy density of a star) always
  holds for a rapid hadron-to-quark conversion, but does not determine
  stability against gravitational collapse if the conversion is slow.

{ The stellar configurations in the extended stability branch are
  stable against all radial perturbations (considering linear
  perturbations). For this reason, their lifetimes are the same as
    those of the "traditional" stable branch. In all the models we
  have considered, the central density of the terminal-mass
  configuration object is less than 3000 MeV/fm$^3$ (see,
  Fig.\ \ref{m_rho} for more details).}

As can be seem from Fig.\ \ref{Fig_new}, the radii of HSs in the
extended stellar branch may differ from the radii of stars made
entirely of hadronic matter by up to $\sim 1$~km. This property,
therefore, could serve as a distinguishing features between both types
of stars.

Another observation to be made from Fig.\ \ref{Fig_new} concerns the
role of the strength of the vector interaction among quarks,
$\eta_v$. As can be seen, increasing the value of $\eta_v$ leads
greater maximum stellar masses, while, at the same time, the extended
branches of the HSs shrink. {{The upper limit on the value of $\eta_v$
    is obtained when the extended branch has shrunk to zero, in which
    case stability ends at the maximum-mass star of the stellar
    sequence. In what follows, we will study two limiting cases for
    $\eta_v$, one where its value is determined by the conventional
    maximum-mass stability criteria mentioned just above (denoted
    $\eta_ {v, {\rm max}}$). The other case corresponds to the minimum
    value of $\eta_v$ (denoted $\eta_{v, {\rm min}}$) determined by
    the requirement that at least $2.05 \, M_\odot$ be obtained with
    our models.  These cases are shown in Fig.\ \ref{MR_hyb} for
    stellar populations with and without $\Delta(1232)$ isobars.  If
    no $\Delta$s are taken into account, the minimum and maximum
    values for $\eta_v$ are $\eta_ {v, {\rm min}} = 0.358$ and $\eta_
    {v, {\rm max}}= 0.471$. If $\Delta$s are taken into account in the
    calculation, we have $\eta_ {v, {\rm min}} = 0.370$ and $\eta_ {v,
      {\rm max}}= 0.483$ for a relative $\sigma$--$\Delta$ coupling of
    $x_{\sigma\Delta}=1.1$, and $\eta_ {v, {\rm min}} = 0.335$ and
    $\eta_ {v, {\rm max}}= 0.470$ for $x_{\sigma\Delta}=1.258$.}}

\begin{figure}[htb]
\begin{center}
\includegraphics[width=0.49\textwidth]{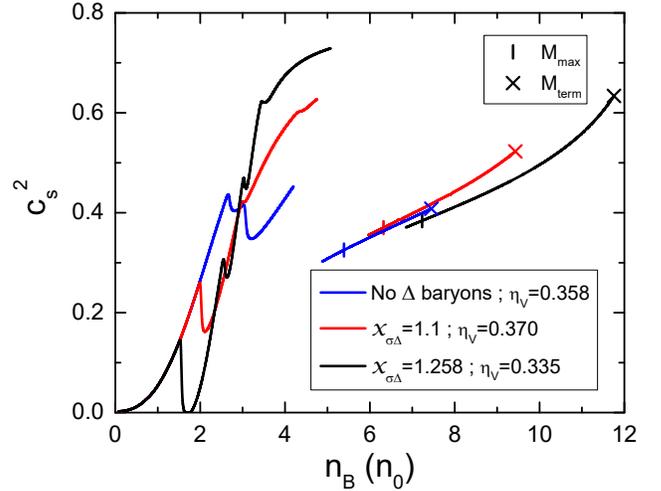}
\caption{(Color online) Square of the speed of sound, $c_s^2$, as
    a function of normalized baryonic number density, $n_B/n_0$, for
    different values of the vector repulsion parameter $\eta_v$.}
 \label{cs_nb}
\end{center}
\end{figure}

In Fig.\ \ref{cs_nb} we show the square of the speed of sound, $c_s^2$, as a
  function of baryon number density for the hybrid star EoS with color
  superconductivity. The locations of the maximum-mass stars are
  marked with vertical bars and the crosses show the stellar models at
  the endpoints of stability.  The erratic behavior of $c_s^2$ below
  around $4\, n_0$ has its origin in the $\Delta$ population (see
  Fig.\ \ref{Pops_Hyb}), which depends on the $\sigma$--$\Delta$
  coupling ratio $x_{\sigma \Delta}$.  A case in point is $x_{\sigma
    \Delta} = 1.258$, for which the $\Delta^-$ population sets
  in at densities even less than $2\, n_0$,  leading to a sharp drop in $c_s^2$.
The zig-zag behavior of $c_s^2$, therefore, is the more prominent the
larger the value of $x_{\sigma \Delta}$.  The speed of sound in the
quark phase, which is present at densities greater than $\sim 5\,
n_0$, violates the so-called conformal limit of $c_s^2 \leq 1/3$ (a
discussion if this limit can be found it
Refs.\ \cite{Bedaque:2014sqa,Tews2018:ApJ}) and reaches values of up
to 0.8 in the cores of HSs at the terminal mass ($M_{\rm term}$). The
actual value of $c_s$ depends, like it is the case for hadronic
matter, on the stiffness of the hybrid EoS which, in turn, is
determined by the value of the strength of the vector interaction,
$\eta_v$. We note that in order to obtain heavy ($\sim 2 \, M_\odot$)
NSs combined with relatively small radii in the 10 to 12 km range
(Fig.\ \ref{MR_hads}), a rapid increase in pressure in the core of a
NS is required. This implies a non-monotonic behavior of the speed of
sound in dense NS matter, which is obtained naturally if the matter in
the cores of NSs is no longer described in terms of nucleons only
\cite{Tews2018:ApJ}.

\subsection{Notes on the mixed hadron-quark phase}

\begin{figure}[htb]
\begin{center}
\includegraphics[width=0.48\textwidth]{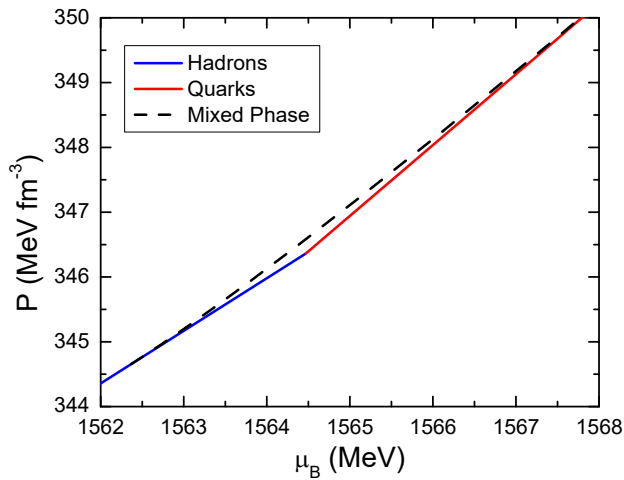}
\caption{(Color online) Pressure as a function of baryon chemical
  potential, for the Maxwell (solid lines) and Gibbs (dashed line)
  construction.}
 \label{EoS_Gibbs}
\end{center}
\end{figure}

\begin{figure}[htb]
\begin{center}
\includegraphics[width=0.48\textwidth]{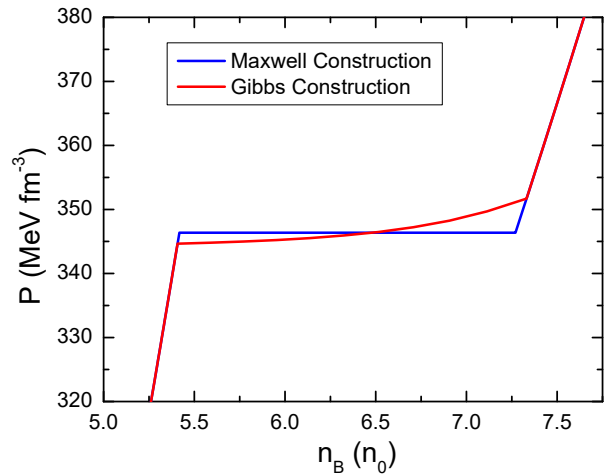}
\caption{(Color online) Pressure as a function of baryon number
  density, in units of the nuclear saturation. The phase transition is
  modeled as both a Maxwell and a Gibbs transition.}
 \label{EoS_comp}
\end{center}
\end{figure}

Even for the smaller values of $\eta_v$ studied in this paper, we
  obtain hybrid stars with only a modest amounts of matter in the
  mixed hadron-quark phase. This feature can be inferred graphically
  from Fig.\ \ref{EoS_Gibbs}, where we show the pressure in the
  quark-hadron transition region obtained for the Gibbs and the
  Maxwell treatment. For the Gibbs phase transition (dashed line) we
find that the mixed phase exists only for baryon chemical potentials
in the small range of $1563~ {\rm MeV} \lesssim \mu_B \lesssim
1568~{\rm MeV}$.  It can also be seen that the phase transition occurs
not until relatively high pressure values are reached.
We note that we have assumed a surface tension between the confined
and deconfined phases of $\sigma_{HQ}=0$ when constructing the mixed
phase and have not taken into account the possibility of structure
formation in the mixed phase
\cite{Glendenning:2001pr,Spinella:2016EJP,Weber2019}.  A comparison of
our results for the EoS shown in Fig.\ \ref{EoS_comp} with those of
Ref.\ \cite{pasta:2019} leads us to conclude that the formation of a
Gibbs mixed phase is not favored by our models and that the phase
transition ought to be Maxwell-like. {{Moreover, it can be seen
      from Fig.\ \ref{EoS_Gibbs} that we obtain a narrow mixed phase region
    for the Gibbs construction of the phase transition. The situation
    is the same for all the hybrid EoS parameters: the
    result is a short mixed phase region of constant pressure inside
    the star with a sharp interface boundary between hadronic and quark
    matter. If $\sigma_{QH} \neq 0$, this
    would suggest that the formation of geometrical structures due to 
    charge rearrangement in the mixed phase is energetically
    disfavored, and that the Gibbs phase transition becomes a Maxwell
    phase transition \cite{pasta:2019}. }}

\begin{figure}[htb]
\begin{center}
\includegraphics[width=0.48\textwidth]{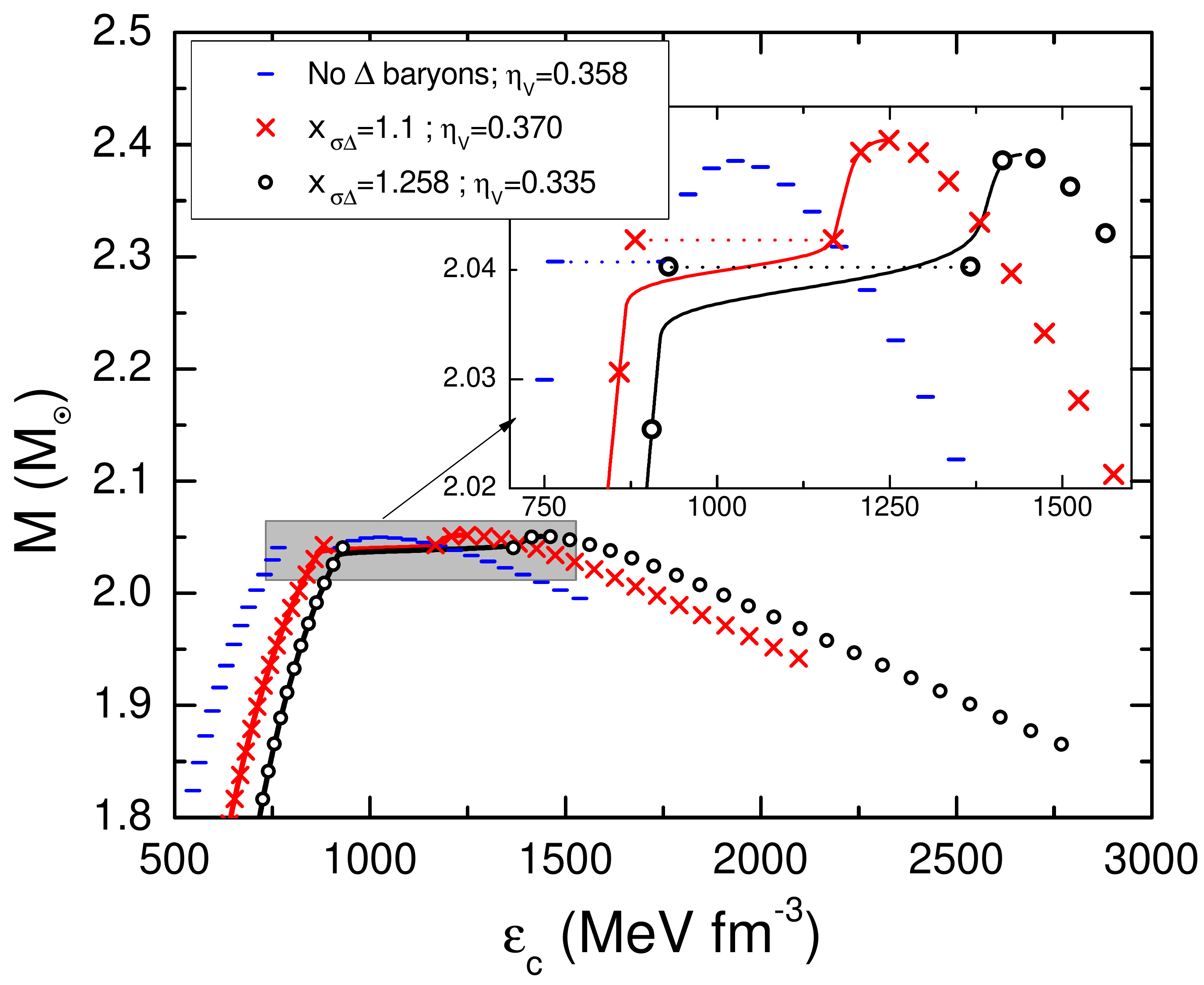}
\caption{(Color online) Gravitational mass as a function of 
  central energy density for the hybrid EoS of this work. The enlarged
  region shows the results obtained for the Maxwell (dotted lines)
  or Gibbs (continuous line) constructions.}
 \label{m_rho}
\end{center}
\end{figure}

The mass-radius relationships obtained for the Gibbs and the Maxwell
treatments are however very similar to each other, except that the
Gibbs stellar sequence terminates at the maximum-mass configuration of
this sequence, while the Maxwell sequence extends stably beyond the
maximum-mass configuration, ending at the terminal mass (see
Sect.\ \ref{ssec:extended_branch}). {{This can be seen also in
    Fig.\ \ref{m_rho}. From the enlarged region in this plot, one
    sees that no mixed phase formation is possible if $\Delta$
    baryons are absent (stellar  configurations marked with blue
    horizontal bars). Both Maxwell and Gibbs constructions are possible for
    the two limiting values of $x_{\sigma\Delta}=1.1$ (red crosses and
    continuous red line) and $x_{\sigma\Delta}=1.258$ (black hollow
    circles and black line). }}

\subsection{Tidal deformability}

A new observational window on the inner workings of NSs is provided by
the gravitational-wave peak frequency and the stellar tidal
deformability of NSs in binary NS mergers
\cite{Bauswein:2019.PRL}. The binary tidal deformability is given by
\begin{equation}
\tilde{\Lambda} = \frac{16}{13}\frac{(12q+1)\Lambda_1 + (12 + q)q^4
  \Lambda_2}{(1+q)^5} \, ,
\end{equation}
where $q=M_2/M_1$ is the ratio of the masses of the merging neutron
stars and $\Lambda _{1,2}$ their dimensionless tidal deformabilities,
which can be calculated by solving an additional differential equation together with the Tolman-Oppenheimer-Volkoff stellar
structure equations \cite{Hinderer2008,PhysRevD.99.083014}.

The tidal deformabilities of the two binary components of GW170817 has
been estimated by the LIGO-VIRGO collaboration
\cite{GW170817-detection}. Although some concerns regarding an
apparent discrepancy between data coming from Handford and Livingstong
LIGO detectors has been recently raised \cite{tidal-discrepancy},
there is agreement on the validity and strength of the general results
obtained by both collaborations. The tidal deformability has been
used to eliminate several EoS that have been used in the past to
describe the matter in the cores of NSs (see
Ref.\ \cite{Orsaria:2019ftf}, and references therein).

\begin{figure}[htb]
\begin{center}
\includegraphics[width=0.48\textwidth]{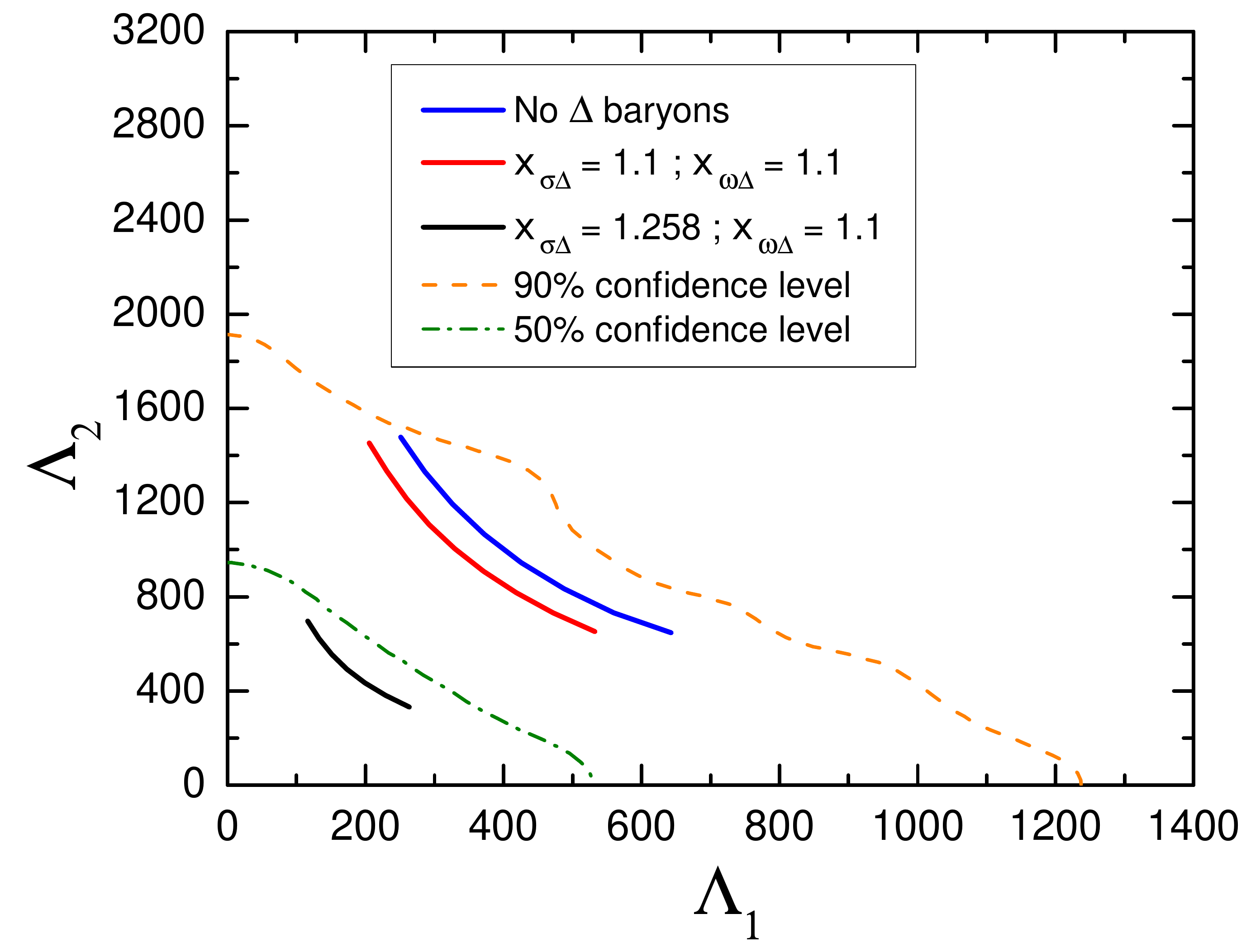}
\caption{(Color online) Dimensionless tidal deformabilities
  $\Lambda_1$ and $\Lambda_2$ for the hadronic configurations of
  Fig.\ \ref{MR_hads}. The orange (cyan) dashed lines represents the
  50\% (90\%) confidence limit of the probability contour of
  GW170817.}
 \label{L1vsL2}
\end{center}
\end{figure}

In Fig.\ \ref{L1vsL2} we present the tidal deformabilities $\Lambda
_1$ and $\Lambda _2$ computed for the NSs shown in
Fig.\ \ref{MR_hads}. The black (gray) curve in this figure denotes the
50\% (90\%) confidence level curve obtained in
Ref.\ \cite{GW170817-detection} for the low-spin scenario. One sees that
the presence of $\Delta$s leads to a better agreement with the
observed data of GW170817. It is not possible to extract any
information related to the appearance of quark matter from the tidal
deformability data of GW170817 as the masses of the objects in this
BNS are lower than the masses at which our models predict the
existence of quark matter in NSs. Event GW190425 involved more massive
NSs \cite{gw190425-detection}. But this event was only observed by the
LIGO Livingston detector and no electromagnetic counterpart was
detected either so that the data from this merger are less
constraining and informative than those of GW170817.

\section{Conclusions}\label{sec:level5} 

In this work we have presented a hybrid EoS which leads to masses which satisfy the latest constraints established by
    massive pulsars and a hadronic EoS satisfying the restrictions on radii set by gravitational-wave data and NICER data.  To
describe matter in the stellar cores of NSs, we have included (in
addition to hyperons) all charged states of the $\Delta(1232)$ baryon
in a non-linear density dependent mean-field treatment based on the
SW4L parametrization and studied the impact of these particles on the
masses and radii of NSs. Specifically the latter depend rather
sensitively on the value of the $\sigma$--$\Delta$ coupling ratio,
$x_{\sigma \Delta} = g_{\sigma\Delta}/g_{\sigma N}$, which has been
taken to be between 1.1 and its maximum-possible value of 1.258 set by
the microscopic stability of the matter. We found that varying
$x_{\sigma\Delta}$ in this range changes the radii of NSs by up to
$\sim 1.5$ km, depending on gravitational mass. The speed of sound,
$c_s$, of hadronic matter remains always less than the speed of light
for $1.1 \leq x_{\sigma \Delta}\leq 1.258$, so that the hadronic EoSs
are causal at all densities. Depending on the value of $x_{\sigma
  \Delta}$, we find that the presence of $\Delta$s in NS matter
drastically alters the speed of sound, which then would no longer be a
monotonically increasing with density, which allows one to accommodate
heavy NSs with relatively small radii in the 10 to 12 km range.

Quark matter is modeled in the framework of the SU(3) non-local NJL
model.  The effects of color superconductivity on the EoS has been
taken into account by considering the 2SC+s diquark condensation
pattern, for the first time, in a non-local NJL
model.  Compared to normal quark matter, the 2SC+s phase is
generally energetically favored over normal quark matter at all
densities. A sequential phase transition from hadronic matter, to
normal quark matter, to 2SC+s quark matter would only be possible if
the hadron-quark phase transition were to occur at the same density at
which normal quark matter turns into 2SC+s matter.  Compared to the
non-color superconducting case, the inclusion of 2SC+s color
superconductivity softens the EoS mildly, which in turn decrease of
the maximum HS mass.  Nevertheless, it is still possible to satisfy
the $2 \, M_\odot$ constraint set by the most massive NSs observed to
date.

We have constructed the phase transition to quark matter using both
the Maxwell and Gibbs descriptions.  If the phase transition is
treated as being sharp (Maxwell), so that no mixed phase exists, two
possible scenarios emerge: either a rapid or a slow phase
transition. As was found in several previous works
\cite{Ranea:2016,Malfatti2019hot}, assuming a slow phase transition
extends the region of stable stars beyond the maximum-mass star of a
given stellar sequence, leading to new stellar configurations that are
more compact than the stars along the traditional branch. The stars on
the extended branch have the same mass as their counterparts on the
traditional branch, but their respective radii differs by up to 1 km
leading to twin-like stellar configurations.  When a rapid phase transition
    is assumed to occur, on the other hand, the extended branch
    vanishes and one is left with only the traditional branch of
    stable configurations. In this case, the appearance of a quark
    matter in the cores of NSs almost immediately destabilizes them
    (aside from a very short portion on the traditional branch), in
    agreement with the results of previous works (see, for example,
    Refs.\ \cite{Ranea:2016,Malfatti2019hot}, and references therein).

As shown recently in Ref.\ \cite{lugones-extended}, treating the
 hadron-quark phase transition as a sharp Maxwell transition leads to
  stable compact stars that, remarkably, lie beyond the maximum-mass
  peak of a stellar sequence (the extended branch). This extended
  branch exists only for the Maxwell transition, but disappears if the
  phase transition is treated as a smooth Gibbs transition. The stars
  on the extended branch contain pure quark matter in their cores, in
  sharp contrast to the stars on the conventional branch which, at
  best, contain only small amounts of quark matter mixed with hadronic
  matter.  The surface tension is known to play a critical role when
  modelling the hadron-quark phase transition in terms of either a
  Maxwell or Gibbs transition. If either one of them has its physical
  correspondence in the core of a compact star, the discussion of this
  paragraph may help to shed light on the unknown value of the surface
  tension $\sigma_{HQ}$ between the confined and deconfined phases \cite{Glendenning:2001pr,Spinella:2016EJP,Weber2019,pasta:2019}.

Possible observable features that may allow one to distinguish between
stars on the conventional branch and stars on the extended branch are
differences in the stellar radii, which could be as large as around 1
km, and the non-radial quasi-normal modes, such as g-modes. As shown,
for instance, in
Refs.\ \cite{g-mode_01,g-mode_02,g-mode_03,Ranea-Sandoval:2018bgu,Orsaria:2019ftf},
g-modes are only present in compact stars if the nuclear EoS contains
a sharp (i.e., constant pressure) density discontinuity.

We also calculated the individual tidal deformabilities $\Lambda_1$
and $\Lambda_2$ of merging NSs for our hadronic EoSs. The
results are consistent with the observational constraints from the
analysis of GW170817 data. This is very interesting as it shows that a
hadronic EoS which includes all particles of the baryonic octet plus
all charged states of the $\Delta(1232)$ is in agreement with present
gravitational-wave data, as well as the latest observed data on masses
and radii of NSs.

\begin{acknowledgments}
  
GM, MO, IFR-S and GAC thank CONICET and UNLP for financial support
under grants PIP-0714 and G140, G157, X824. IFR-S is thankful for
hospitality extended to him at the San Diego State University and for
the support received from the Fulbright Foundation and CONICET via a
Fulbright/CONICET joint scholarship.  FW is supported through the
U.S.\ National Science Foundation under Grants PHY-1714068 and
PHY-2012152.
\end{acknowledgments}

\appendix

\section{\label{append}Details on the treatment of the 2SC+s phase}

Working in SU(3), we can write the operator $A(p)$ of
Eq.\ (\ref{action}) in a compact form,
\begin{equation}
A(p) = 
 \begin{pmatrix}
-\slashed{p} + \hat{M} + i\hat{\mu}\gamma^4 & i\sum_A \Delta_A^p
\gamma_5 \lambda_A \lambda_A \\ i\sum_A (\Delta_A^p)^* \gamma_5
\lambda_A \lambda_A& -\slashed{p} + \hat{M} - i\hat{\mu}\gamma^4  \\
\end{pmatrix}\,  ,
\label{operator}
\end{equation}
where $\Delta_A^p = \Delta_A g(p)$. This operator is a 72 $\times$ 72
matrix in Dirac, flavor, color and Nambu-Gorkov spaces. However, it is
possible to evaluate the trace in Eq.\ (\ref{action}). Note the matrix
of Eq.\ (\ref{operator}) is the inverse fermion propagator, where
$\hat{M} = \mathrm{diag}(M_{u}, M_{d}, M_{s})$
\cite{Buballa:2005}. Then, rearranging rows and columns, and using the
logarithm trace notation we can write
\begin{eqnarray}
\label{matricessupercond}
 \mathrm{Tr}\{\mathrm{ln}[A(p)]\} =&& \mathrm{Tr}[\mathrm{ln}
   (M_{ug,dr})] + \mathrm{Tr}[\mathrm{ln} (M_{ur,dg})]
 \nonumber\\ &+&\mathrm{Tr}[\mathrm{ln} (M_{ub,sr})] +
 \mathrm{Tr}[\mathrm{ln} (M_{ur,sb})] \nonumber \\ &+&
 \mathrm{Tr}[\mathrm{ln} (M_{db,sg})]+ \mathrm{Tr}[\mathrm{ln}
   (M_{dg,sb})] \nonumber\\ &+& \mathrm{Tr}[\mathrm{ln}
   (M_{ur,dg,sb})]   \, .
\end{eqnarray}

In the framework of the 2SC+s phase, $\Delta_2 \neq 0$ and $\Delta_5 =
\Delta_7 = 0$. Thus, the matrices $M_{f,f'}$ of
Eq.\ (\ref{matricessupercond}) involving the quark strange do not have
diagonal components and such quark decouples. Finally, the only matrix
structure involving diquarks in a compact form is given by
\begin{equation}
M_{ud} = 
 \begin{pmatrix}
-\slashed{p} + \hat{M} + i\hat{\mu}\gamma^4 & i\Delta_2^p \gamma^5
\\ i\Delta_2^p \gamma^5 & -\slashed{p} + \hat{M} - i\hat{\mu} \gamma^4
\\
\end{pmatrix} \, ,
\end{equation}
simplifying the problem to calculate now the determinant of
$M_{ud}$. By adding the decoupled part due the presence of the strange
quark we obtain
\begin{equation}
\mathrm{Tr} [\mathrm{ln}\,A(p)]=\nonumber\\
\sum_c  \mathrm{ln}(q_{sc}^{+\,2} + M_{sc}^2)+ \frac{1}{2} \mathrm{ln} |A_c|^2 \, ,
\end{equation}
where
\begin{eqnarray}
 A_c &=& \left[{q_{uc}^+}^2 + M_{uc}^2\right]\left[{q_{dc}^-}^2 +
   {M_{dc}^*}^2\right] \nonumber\\ && \left(1 -
 \delta_{bc}\right){\Delta^p}^2\left[{\Delta^p}^2 + 2 q_{uc}^+
   . q_{dc}^- + 2M_{uc}M_{dc}^* \right], \nonumber\\ q_{fc}^\pm &=&
 \left( p_0 \mp i \left[\mu_{fc} - \bar \theta_f g\left({p_{fc}^\pm}^2
   \right) \right], \bm{p} \right), \nonumber \\ p_{fc}^\pm &=& \left(
 p_0 \mp i \mu_{fc}, \bm{p} \right), \nonumber \\ M_{fc} &=& m_f +
 \bar \sigma_f \, g\left( {p_{fc}^+}^2 \right) , \nonumber\\ \Delta^p
 &=& {\bar \Delta}\, \tilde{g} \, , \nonumber
\end{eqnarray}
being
\begin{equation}
\tilde{g}=g\left(\frac{ \left[{p_{ur}^+} + {p_{dr}^-} \right]^2
}{4}\right).\nonumber
\end{equation}
The potential of Eq.\ (\ref{Omegon}) is regularized to avoid
divergences for finite values of the current quark mass. The
regularization procedure can be expressed through the relation
\begin{eqnarray}
\Omega=\Omega^{MFA}-\Omega^{free}+ \Omega_{free}^{Reg} \, ,
\end{eqnarray}
which is equivalent to Eq.\ (\ref{Omegon}), and where $\Omega^{free}$
is obtained by setting $\bar{\sigma}=\bar\theta=\bar\Delta=0$ and
\begin{eqnarray} 
 \Omega_{free}^{Reg}=\sum_{f,c} \frac{1}{24\pi^2}\Biggl[\left(-5m_f^2
   + 2\mu_{fc}^2\right)\mu_{fc}\sqrt{\mu_{fc}^2 - m_f^2} \nonumber
   \\ + 3m_f^4\mathrm{ln}\left(\frac{\mu_{fc} + \sqrt{\mu_{fc}^2 -
       m_f^2}}{m_f}\right) \Biggr] \Theta\left(\mu_{fc} -
 m_f\right) \, , \nonumber
\end{eqnarray}
is the regularized thermodynamic potential of a free fermion gas.

To compute the auxiliary fields $S_f$, $V_\theta$ and $D$ we use that
$\frac{\partial {\rm ln} |A_c|^2}{\partial k} = \frac{\partial \rm{ln}
  |A_c^* A_c|}{\partial k} = 2{\rm Re}\left(\frac{1}{A_c}
\frac{\partial A_c}{\partial k} \right)$. Thus, for quarks $u$ and $d$
the auxiliary fields in Eq.\ (\ref{Omegon}) associated with the
mean-fields $\bar\sigma_u$ and $\bar\sigma_d$ can be written as
\begin{eqnarray}
  \bar S_f = -4\sum_{c = r,g,b}
  \int_{0}^{+\infty}dp_0\int_{0}^{+\infty} \frac{p^2\, dp}{\pi^3}\,
  \mathrm{Re}\,\left[\frac{B_{fc}}{A_c}\right] \, ,
\end{eqnarray}
where
\begin{eqnarray*}
B_{uc}&=&M_{uc} \,g( p_{uc}^{+2} ) \left[q_{dc}^{-2} + M_{dc}^{*2}
  \right] + g(p_{uc}^{+2}) \,(1 - \delta_{bc})\,
\Delta^{p2}\,M_{dc}^*,\nonumber\\ B_{dc}&=&M_{dc}^*
g^*(p_{dc}^{+2})\left[q_{uc}^{+2} + M_{uc}^2\right]+ g^*(p_{dc}^{+2})
\,(1 - \delta_{bc}){\Delta^p}^2\,M_{uc}.
\end{eqnarray*}
The auxiliary field associated to $\bar\sigma_s$ reads
\begin{equation}
\bar S_s = -4\sum_{c = r,g,b}\int_{0}^{+\infty}dp_0\int_{0}^{+\infty}
\frac{p^2 \,dp\,}{\pi^3} \mathrm{Re}\left[\frac{M_{sc}
    \,g(p_{sc}^{+2}) }{q_{s}^{+2} + M_{sc}^2}\right].
\end{equation}
For the auxiliary field associated to $\bar\theta_u$, $\bar\theta_d$
and $\bar\theta_s$ we have
\begin{equation}
  \bar V_f = -4\,\sum_{c = r,g,b}
  \int_{0}^{+\infty}dp_0\int_{0}^{+\infty} \frac{p^2\,dp}{\pi^3} \,
  \mathrm{Re}\left[\frac{C_{fc} }{A_c}\right] ,
\end{equation}
where
\begin{eqnarray}
C_{uc}&=&i\,q_{0uc}\,g(p_{uc}^{+2})\left[q_{dc}^{-2} + M_{dc}^{*2}\right]\nonumber\\
&+& i\,g(p_{uc}^{+2})\,(1 - \delta_{bc})\,{\Delta^p}^2\,q_{0dc}^-  \,  , \nonumber\\
C_{dc}&=&-i\,q_{0dc}\,g(p_{dc}^{-^2})\left[q_{uc}^{+2} + M_{uc}^2\right]  \,  , \nonumber\\ 
& -& i\,g(p_{dc}^{-2})\,(1 - \delta_{bc})\,{\Delta^p}^2\,q_{0uc}^+ \,  , 
\end{eqnarray}
and
\begin{equation}
\bar V_s = -8\sum_{c =
  r,g,b}\,\int_{0}^{+\infty}dp_0\int_{0}^{+\infty} \frac{p^2
  \,dp\,}{\pi^3} \mathrm{Re}\left[\frac{i\,q_{0s}\,g\left({p_s^+}^2
    \right) }{q_s^2 + M_s^2}\right] ,
\end{equation}
where $q_{0fc}^\pm$ is the zeroth component of $q_{fc}^\pm$  . Finally, the auxiliary field
related with the mean-field $\bar\Delta$ is given by
\begin{equation}
 \bar{D} = -2\sum_{c = r,g} \int_{0}^{+\infty}dp_0\int_{0}^{+\infty}
 \frac{p^2 \,dp\,}{\pi^3} \mathrm{Re}\left[\frac{D_{ud}}{A_c}\right]  ,
\end{equation}
where
\begin{eqnarray*}
D_{ud}=2{\Delta^p}^3 \tilde{g} + {\Delta^p} \tilde{g}\,\left(2
q_{uc}^+ . q_{dc}^- + 2M_{uc}M_{dc}^*\right) .
\end{eqnarray*}

%


\end{document}